\documentclass[12pt]{article}

\usepackage{amsmath,amssymb,psfig,latexsym,theorem}
\usepackage[all]{xy}

\begin{document}

\hfill ILL-(TH)-02-10

\hfill hep-th/0212218

\vspace{0.5in}

\begin{center}

{\large\bf D-branes, Orbifolds, and Ext groups }

\vspace{0.25in}

Sheldon Katz$^{1,2}$, Tony Pantev$^3$, and Eric Sharpe$^{1}$ \\
$^1$ Department of Mathematics \\
1409 W. Green St., MC-382 \\
University of Illinois \\
Urbana, IL  61801 \\
$^2$ Department of Physics\\
University of Illinois at Urbana-Champaign\\
Urbana, IL 61801\\
$^3$ Department of Mathematics \\
University of Pennsylvania \\
David Rittenhouse Lab. \\
209 South 33rd Street \\
Philadelphia, PA  19104-6395 \\
{\tt katz@math.uiuc.edu},
{\tt tpantev@math.upenn.edu},
{\tt ersharpe@uiuc.edu} \\

 $\,$

$\,$

\end{center}

In this note we extend previous work on massless Ramond spectra of
open strings connecting D-branes wrapped on complex manifolds, to
consider D-branes wrapped on smooth complex orbifolds.  Using standard
methods, we calculate the massless boundary Ramond sector spectra
directly in BCFT, and find that the states in the spectrum are counted
by Ext groups on quotient stacks (which provide a notion of
homological algebra relevant for orbifolds).  Subtleties that cropped
up in our previous work also appear here.  We also 
use
the McKay correspondence to relate Ext groups on quotient stacks to
Ext groups on (large radius) resolutions of the quotients.
As stacks
are not commonly used in the physics community, we include pedagogical
discussions of some basic relevant properties of stacks.

\begin{flushleft}
December 2002
\end{flushleft}

\newpage

\tableofcontents

\newpage

\section{Introduction}

One of the predictions of current proposals for the physical
significance of derived categories (see
\cite{paulron,medc,mikedc,paulalb} for an incomplete list of early
references) is that massless boundary Ramond sector states of open
strings connecting D-branes wrapped on complex submanifolds of
Calabi-Yau's, with holomorphic gauge bundles, should be counted by
mathematical objects known as Ext groups.  Specifically, if we have
D-branes wrapped on the complex submanifolds $i: S \hookrightarrow X$
and $j: T \hookrightarrow X$ of a Calabi-Yau $X$, with holomorphic
vector bundles ${\cal E}$, ${\cal F}$, respectively, then massless
boundary Ramond sector states should be in one-to-one correspondence
with elements of either
\begin{displaymath}
\mbox{Ext}^n_X\left( i_* {\cal E}, j_* {\cal F} \right)
\end{displaymath}
or
\begin{displaymath}
\mbox{Ext}^n_X\left( j_* {\cal F}, i_* {\cal E} \right)
\end{displaymath}
(depending upon the open string orientation).

In \cite{orig} this proposal was checked explicitly
for large radius Calabi-Yau's
by using standard well-known BCFT methods to compute 
the massless Ramond sector spectrum of open strings connecting
such D-branes, and then relating that spectrum to Ext groups.
Ext groups can be related to ordinary sheaf cohomology groups 
via spectral sequences.
In the absence of background gauge fields, BRST invariance of a
vertex operator dictates that the corresponding bundle-valued
differential form is $\bar\partial$-closed, leading to sheaf 
cohomology groups.  When background gauge fields are turned on,
the BRST-invariance condition can be decomposed into 
tangential and normal components.  The tangential components 
again lead to $\bar\partial$-closed
differential forms.  The normal components can be related to the tangential
components by the boundary conditions as modified by the background
gauge fields.
At least in cases we understand well, these conditions can be reinterpreted
as the vanishing of differentials in the
spectral sequences we mentioned above.  The result is that
the massless Ramond sector spectrum is counted directly by Ext groups.

There are numerous possible followups to the work in \cite{orig}.
In this paper we shall extend the methods of \cite{orig}
to the case of string orbifolds.  In particular,
we shall demonstrate that there is a one-to-one correspondence
between massless boundary Ramond sector states in open string
orbifolds and either
\begin{displaymath}
\mbox{Ext}^n_{ [X/G] }\left( i_* {\cal E}, j_* {\cal F} \right)
\end{displaymath}
or
\begin{displaymath}
\mbox{Ext}^n_{ [X/G] }\left( j_* {\cal F}, i_* {\cal E} \right)
\end{displaymath}
(depending upon the open string orientation),
where $[X/G]$ is known as a ``quotient stack.''
We do not make any assumptions about the physical relevance
of stacks; rather, we compute the massless boundary Ramond
sector spectrum directly in BCFT, from first-principles,
and then, after some work, notice that said spectrum happens
to be counted by Ext groups on quotient stacks.

Our work allows us to
\begin{itemize}
\item Extend calculations of massless open string spectra in orbifolds
to more general D-brane configurations than previously
calculable.
\item Gain another handle on the extent to which derived categories
are relevant to branes, by not only describing a notion of Ext groups
relevant away
from large radius limits, but by calculating that
that notion of Ext groups (on quotient stacks) is physically
relevant and counts massless boundary Ramond sector open string states.
\item Gain insight into the physical relationship between
string orbifolds and quotient stacks.
\end{itemize}

In principle, calculating the massless boundary Ramond sector
spectrum in an open string orbifold is straightforward,
following methods introduced in \cite{dougmoore}.
The worldvolumes of the D-branes must be closed under the G-action,
and
one must specify an orbifold group action on the Chan-Paton factors\footnote{
More generally, whenever one has a field with a gauge symmetry,
one must specify the orbifold group action on the field, not just
the base space.  Specifying an action on the base space is insufficient
to determine the orbifold of the theory, as the orbifold group
action on a field with a gauge symmetry can be combined with a gauge
transformation to get a new orbifold group action.
For example, in the case of heterotic orbifolds,
the possible choices of orbifold group action on the heterotic
gauge field are known as orbifold Wilson lines.
The possible orbifold group actions on the B field
include discrete torsion \cite{medt,dtrev}.}
in addition to the orbifold group action on the base space.
Once one has specified those orbifold group actions,
the massless spectrum in the open string orbifold is simply the
$G$-invariant part of the massless open string spectrum on the
covering space.  There are no twist fields in the usual sense
of closed string orbifolds.
The
resulting spectrum is counted by a mathematically sensible notion of
Ext groups for orbifolds, specifically, Ext groups on quotient stacks,
as mentioned above.

For readers not familiar with the terminology,
quotient stacks are a mathematical device for handling orbifolds.
More generally, stacks\footnote{In this paper, when we refer
to stacks, we shall be referring to the $\mathrm{C}^\infty$ counterpart
of smooth Deligne-Mumford stacks.  These are examples of what are properly
called \'etale groupoids.}
are the next best thing to spaces.  One can do differential geometry
on stacks; one can define bundles and sheaves on stacks.
In practice, most mathematical notions that physicists are familiar
with from spaces, carry over to stacks.

Quotient stacks (denoted with brackets as $[X/G]$ to distinguish
them from quotient spaces $X/G$)
also have the property of encoding orbifold data.
For example, a coherent sheaf on a quotient stack $[X/G]$
is the same thing as a sheaf on $X$ together with a $G$-equivariant
structure (that is, a choice of $G$-action on the sheaf).
In general, $G$-equivariant sheaves on a cover are not quite
the same thing as sheaves on a quotient space, as we shall see
later, so this distinction is not purely academic.

In the special case that $G$ acts freely on $X$, so that
the quotient space $X/G$ is smooth, the quotient stack $[X/G]$
is the same as the quotient space $X/G$.\footnote{All smooth algebraic
varieties are smooth Deligne-Mumford stacks.  Stacks generalize
spaces, in this sense.}  So, in the case that
$G$ acts freely, our results outlined above reproduce the
results of \cite{orig}, in that we recover the classification of
massless Ramond sector states on large radius Calabi-Yau's by Ext groups
on the Calabi-Yau's.  Quotient stacks differ from quotient spaces
when $G$ does not act freely; in a sense, quotient stacks encode
additional information over the fixed points of the group action.

Now, it has also been suggested
that not only might string propagation
on stacks be a sensible notion, but string orbifolds are the same
as strings propagating on quotient stacks.
One consequence of such a claim, together with recent proposals
on derived categories in physics, is that the massless boundary Ramond
sector states in open strings in orbifolds should be counted by Ext
groups on stacks.  We are checking that statement in this paper,
and therefore providing evidence for such a proposal.
In previous work on this subject, a proposal has been made
for a generalization of the classical action of sigma models on
spaces to include stacks \cite{meqs}.  
For example, formally a sigma model on a stack has
twisted sectors in a very natural sense.
The same paper outlined how
many physical characteristics of string orbifold CFT's could
have a purely geometric understanding in terms of stacks.
For example, a quotient stack $[X/G]$ is smooth so long as
$X$ is smooth and $G$ acts by diffeomorphisms,
regardless of whether or not $G$ has fixed points.
The sense in which a stack is smooth naturally ties into
the well-behavedness of string orbifold CFT's.
In other work, it has been shown that certain correlation functions
in orbifold CFT's can be rewritten in a form involving stacks
that is analogous to correlation functions in sigma models on
ordinary spaces \cite{agv}.  Much more work remains to be
done to completely understand the physical relevance of quotient
stacks, including perhaps most importantly, an understanding of
how deformation theory of stacks is consistent with CFT moduli spaces.

We begin in section~\ref{warmup} with a very basic discussion
of quotient stacks and Ext groups thereon, as relevant to the
rest of the paper.  

In section~\ref{parcoin} we begin with
the case of parallel coincident branes.  We describe the
first-principles computation of the massless boundary Ramond sector
spectrum, and relate that spectrum to Ext groups on quotient stacks.
Just as in \cite{orig}, spectral sequences play a key role, and we
describe explicitly how those spectral sequences are realized
physically in BRST cohomology.  We check the spectrum calculation in
numerous examples, which also allow us to show how to use this
technology in a variety of cases.  We also discuss an example in which
the spectral sequence is necessarily nontrivial.  Finally, fractional
branes are examples of D-branes which cannot be understood in terms
of sheaves on quotient spaces; to describe them with sheaves,
one must use quotient stacks, not quotient spaces.  We discuss fractional
branes in some detail.

In section~\ref{pardiff}, we consider parallel branes of different dimensions.
We again calculate massless boundary Ramond sector spectra, and relate those
spectra to Ext groups on quotient stacks.
We check our results by comparing to the ADHM construction on ALE spaces,
as realized in terms of branes on branes.

In section~\ref{gencase} we finally consider the general case, and relate
massless boundary Ramond sector spectra to Ext groups on quotient stacks.

In section~\ref{mckay} we use the McKay correspondence to relate
Ext groups on quotient stacks to Ext groups on resolutions of
quotient spaces.  In other words, orbifold D-brane spectrum computations
can be related to large radius spectrum computations, thanks to the McKay
correspondence.

In appendix~\ref{stxap} we give a slightly more detailed overview
of topological and differential-geometric features of stacks,
than was given in section~\ref{warmup}.

In appendix~\ref{pfs} we describe proofs of the spectral sequences
used in this paper.

Throughout this paper, we shall assume that the orbifold group $G$ is
finite.  Furthermore, we shall be considering orbifolds of large
radius Calabi-Yau's.  In other words, before performing the orbifold
operation, the Calabi-Yau's are at large radius and standard
Born-Oppenheimer-type methods can be used to calculate massless Ramond
spectra.  After the orbifold operation, of course, the resulting
conformal field theory will no longer be entirely at large radius, but
that fact is irrelevant for our analysis.  We shall assume that the
$B$ field is identically zero.  Nonzero flat $B$ fields are also very
interesting, and will form the subject of later work.
We shall also assume that the
large radius Calabi-Yau and all the submanifolds on which we wrap
branes are smooth, and that the finite group $G$ acts by
diffeomorphisms.

\section{Warmup:  sheaves on stacks}  \label{warmup}

Before discussing the massless Ramond sector spectrum calculation,
we shall discuss sheaves and homological algebra on quotient stacks,
as the relevant notions are unlikely to be familiar to many
physicists.
Our discussion of stacks in this section is extremely brief;
a slightly longer exposition can be found in appendix~\ref{stxap}.

As mentioned previously, one can define bundles and sheaves
on stacks, in addition to doing differential geometry on stacks;
indeed, most mathematical notions relevant to spaces carry over
to stacks.  In this paper, we will not need to use any particularly
deep technical notions or definitions; instead, we shall merely
list some basic results, which will be used repeatedly.
For a pedagogical discussion of stacks oriented towards physicists,
see \cite{meqs}; see also \cite{vistoli,gomez,cm,fantechi,lmb}.

In a nutshell, a quotient stack looks like a quotient space
away from the fixed loci; over the
images of the fixed points in  the quotient space, the quotient stack has
additional structure\footnote{There is a map $[X/G]\to X/G$ which
is not an isomorphism in general.  If $Z\subset X$ is the set of all elements
of $X$ fixed by a nonzero element of $G$, then the open substack 
$[(X-Z)/G]$ is the same as the quotient space $(X-Z)/G$. The closed substack
$[Z/G]$ of $[X/G]$ is not isomorphic to the subspace $Z/G$ of $X/G$; this
assertion gives a precise meaning to our statement that there is ``additional
structure'' over the fixed locus.}.
In the case when $G$ acts freely on $X$,
the quotient stack and the quotient space are identical.

When one takes definitions of functions, bundles, {\it etc} on a
general stack and applies them to quotient stacks, one finds that
functions, bundles, {\it etc} on quotient stacks $[X/G]$ are the same
as $G$-equivariant objects on $X$. 
In particular, a coherent
sheaf on a quotient stack $[X/G]$ is the same thing as a coherent
sheaf on $X$ together with a $G$-equivariant structure. 
As a result, given a $G$-equivariant sheaf ${\cal F}$ on $X$, we
will typically abuse notation by using the same symbol ${\cal F}$ to
denote the corresponding sheaf on $[X/G]$. 

In passing, it is important to mention that a
$G$-equivariant sheaf on $X$ is not the same thing as a sheaf on the
quotient space $X/G$. In fact, the sheaves on $X/G$ are precisely
those $G$-equivariant sheaves for which the stabilizer $G_{x}$ of any
point $x \in X$ acts trivially on the stalk of the sheaf at $x$.  This
will be important later when we study fractional branes, which
(immediately from their definition) are $G$-equivariant sheaves on
$X$, and hence sheaves on the stack $[X/G]$, but which we shall see
cannot be understood as sheaves on the quotient space $X/G$ in
general. 

We can manipulate coherent sheaves on stacks in much
the same way that we can manipulate coherent sheaves on ordinary
spaces.  For example, there is a well-defined notion of
homological algebra on stacks.  Consider quotient stacks
for simplicity.  If ${\cal E}$, ${\cal F}$ are
two coherent sheaves on $X$ with $G$-equivariant structures,
or equivalently, two coherent sheaves on the stack $[X/G]$,
then the groups
\begin{displaymath}
\mbox{Ext}^n_{ [X/G] } \left( {\cal E}, {\cal F} \right)
\end{displaymath}
are well-defined.
(Note we have abused notation, using 
${\cal E}$, ${\cal F}$ to
denote the corresponding sheaves on $[X/G]$ as well as the original sheaves
on $X$, as we warned the reader previously.)

Computing Ext groups on quotient stacks is far easier than one might
have naively thought. They are defined as the Ext groups in the
abelian category of coherent sheaves on $[X/G]$, which as we mentioned
above, can be identified with the category $\mathrm{Coh}_{G}(X)$ of
$G$-equivariant coherent sheaves on $X$ (see
\cite[Example~12.4.6]{lmb}).

In general, if ${\cal E}$, ${\cal F}$ are $G$-equivariant sheaves,
then the Ext groups $\mbox{Ext}^n_X({\cal E}, {\cal F})$
on the covering space $X$ form a representation of $G$,
and so have a so-called ``isotypic decomposition'' into
irreducible representations of $G$, as
\begin{displaymath}
\mbox{Ext}^n_X\left( {\cal E}, {\cal F} \right) \: = \:
\bigoplus_{\rho} \mbox{Ext}^n_{X, \rho} \left( {\cal E},
{\cal F} \right).
\end{displaymath}
We shall denote the $G$-invariant part of these Ext groups
with a superscript $G$ as
\begin{displaymath}
\mbox{Ext}^n_X\left( {\cal E}, {\cal F} \right)^G \: = \:
\mbox{Ext}^n_{X, 1} \left( {\cal E}, {\cal F} \right)
\end{displaymath}
Finally, we are ready to discuss how one computes Ext groups
on quotient stacks.  So long as $G$ is a finite group,
the Ext group on the quotient stack is simply the $G$-invariant 
part  
of the Ext group on the covering space $X$: 
\begin{displaymath}
\mbox{Ext}^n_{ [X/G] } \left( {\cal E}, {\cal F} \right) \: = \:
\mbox{Ext}^n_{\mathrm{Coh}_{G}(X)} \left( {\cal E}, {\cal F} \right) =
\mbox{Ext}^n_X \left( {\cal E}, {\cal F} \right)^G 
\end{displaymath}
The last equality follows from the
Grothendieck spectral sequence (see \cite[Section~5.6]{groth}):
$H^{p}(G,\mbox{Ext}^{q}_{X}({\cal E},{\cal F})) \Rightarrow 
\mbox{Ext}^{p+q}_{\mathrm{Coh}_{G}(X)}({\cal E},{\cal F})$ 
and the observation that since $G$ is reductive and the coefficients
are vector spaces, $G$ 
has no cohomology with finite dimensional coefficients of degree $p > 0$.

One can now see the intuition behind the conclusion of
this paper, namely
that massless Ramond states in open string orbifolds
are necessarily computed by Ext groups on quotient stacks.
After all, we argued in \cite{orig} that massless Ramond states
on smooth large radius Calabi-Yau's are counted by Ext groups
on the Calabi-Yau's, and in \cite{dougmoore} it was demonstrated
that one merely takes $G$-invariants to get orbifold spectra.
Since the $G$-invariants of the Ext groups on the cover
are the Ext groups on the quotient stack, surely that
means that the orbifold spectrum is calculated by Ext groups
on the quotient stack.
This argument is fine for gaining intuition,
but it is a little too loose to completely confirm the result.
After all, in \cite{orig}, Ext groups emerged nontrivially,
only as the end-product of a physically-realized spectral
sequence that started with certain sheaf cohomology groups,
so simply taking $G$-invariants 
of Ext groups might miss subtleties if the spectral sequences
are not well-behaved with respect to the group action.
In this paper we shall work much more systematically, repeating the analysis
of \cite{orig} in detail, checking that the spectral sequences
behave well with respect to group actions,
and will eventually find Ext groups on quotient stacks
after a thorough and detailed study.

Closely related results apply to sheaf cohomology.
For any space $X$ with a $G$ action and $G$-equivariant bundle
${\cal E}$ on $S$, the sheaf cohomology groups
$H^n(S, {\cal E})$ each form a representation of $G$, and so
have an ``isotypic decomposition'' into irreducible representations
of $G$:
\begin{displaymath}
H^n\left(S, {\cal E} \right) \: = \:
\bigoplus_{\rho} H^n_{\rho}\left(S, {\cal E} \right)
\end{displaymath}
just as we saw for Ext groups.
We shall denote the $G$-invariant part of sheaf cohomology
with a superscript, as
\begin{displaymath}
H^n\left(S, {\cal E} \right)^G.
\end{displaymath}

For future reference we examine the action of Serre duality in the case
of a quotient stack.
First, if $X$ is a Calabi-Yau and $G$ is a finite group
preserving\footnote{If $X$ were Calabi-Yau but $G$ did {\it not} preserve
the holomorphic volume form, then Serre duality would have
almost the same form, except that the dualizing sheaf would be
${\cal O}_X$ with a {\it non}trivial $G$-equivariant structure.
Although ${\cal O}_X$ is trivial, the change in $G$-equivariant
structure would alter the $G$-invariants.  We shall see an example of
this phenomenon in section~\ref{czn}.} the holomorphic
volume form, then Serre duality acts on Ext groups on quotient
stacks in the fashion one would expect naively from experience
with Ext groups on ordinary Calabi-Yau spaces:
\begin{displaymath}
\mbox{Ext}^p_{ [X/G] }\left( {\cal S}_1, {\cal S}_2 \right)
\: \cong \:
\mbox{Ext}^{n-p}_{ [X/G] } \left( {\cal S}_2, {\cal S}_1 \right)^\vee
\end{displaymath}
where $n = \mbox{dim } X = \mbox{dim }[X/G]$. Furthermore, in terms of
the sheaf cohomology groups, Serre duality relates the isotypic
decomposition of the sheaf cohomology corresponding to dual
irreducible representations of $G$.  In other words, for any
irreducible representation $\rho$ of $G$,
\begin{displaymath}
H^p_{\rho}\left(S, {\cal E}\right)
\: \cong \:
H^{s-p}_{\rho^*}\left(S, {\cal E}^{\vee} \otimes K_S \right)^\vee
\end{displaymath}
where $s = \mbox{dim }S$ and $K_S$ denotes the
anticanonical bundle on $S$.
Thus, in particular, Serre duality exchanges the $G$-invariant
parts of the sheaf cohomology groups.

\section{Parallel coincident branes on $S \hookrightarrow X$}  \label{parcoin}

In this section we shall consider orbifolds of
theories describing parallel coincident D-branes
wrapped on smooth submanifolds $S \hookrightarrow X$,
with holomorphic gauge fields.

\subsection{Basic analysis}

Following \cite{dougmoore}, to study D-branes in orbifolds,
one constructs a set of D-branes on the covering space,
and specifies\footnote{Specifying a group action on the space
does not suffice to specify a group action on a physical
theory containing fields with gauge invariances.  For example,
if one has a bundle or gauge field, the group action on that
bundle or gauge field is not uniquely determined by the group action
on the base, as one can combine any one group action with a set of
gauge transformations.  Such a choice of group action on a bundle,
or gauge field, or coherent sheaf, or other object where
there is ambiguity, is known as a $G$-equivariant structure.} 
$G$-equivariant structures.  Open string spectra are calculated
\cite{dougmoore} by taking $G$-invariants of the spectra on the
covering space.

Thus, computing massless boundary Ramond spectra in open strings
describing D-branes in orbifolds is straightforward,
given the analysis in \cite{orig}.
In principle, we specify a set of D-branes on the covering space,
together with group actions ({\it i.e.} $G$-equivariant
structures) on the Chan-Paton factors.
The spectrum of the orbifold theory then consists \cite{dougmoore}
of the $G$-invariants of the theory on the cover.
There are no twisted sector states in the usual sense of closed
string theory, or rather, the twisted sector states are realized
on the cover as open strings between D-branes corresponding to different
pre-images of a D-brane on the quotient.

The analysis of \cite{orig} assumed the Calabi-Yau was at
large radius, so that the sigma model is weakly coupled.  
Obviously, an orbifold is not at large radius;
however, so long as the covering space theory is at large radius,
the methods of \cite{orig} still apply.
(This trick of using Born-Oppenheimer methods to compute spectra
in orbifolds of large radius theories is standard in
closed-string orbifold spectrum computations.) 

Thus, much of the analysis proceeds as in \cite{orig}.
Let $X$ be a Calabi-Yau, with an action of a discrete group $G$.
Let $S$ be a smooth submanifold that is preserved by $G$,
{\it i.e.}, $G: S \mapsto S$.  Let ${\cal E}$, ${\cal F}$ be a pair
of holomorphic vector bundles on $S$ that are $G$-equivariant, and pick
specific $G$-equivariant structures on both.

As discussed in \cite{orig}, massless boundary Ramond sector states
on the covering space are of the form
\begin{equation}    \label{ramond1}
b^{\alpha \beta j_1 \cdots j_m}_{\overline{\imath}_1 \cdots 
\overline{\imath}_n}(\phi_0) \, \eta^{ \overline{\imath}_1} \cdots
\eta^{ \overline{\imath}_n} \theta_{j_1} \cdots \theta_{j_m}
\end{equation}
where
\begin{eqnarray*}
\eta^{\overline{\imath}} & = & \psi_+^{\overline{\imath}} 
\: + \:
\psi_-^{\overline{\imath}} \\
\theta_i & = & g_{i \overline{\jmath}} \left(
\psi_+^{\overline{\jmath}} \: - \: \psi_-^{\overline{\jmath}} \right)
\end{eqnarray*}
are BRST invariants in the standard notation for the $B$ model \cite{edtft}.
The indices $\alpha$, $\beta$ are Chan-Paton indices, describing
the holomorphic bundles ${\cal E}$, ${\cal F}$.  

The fields $\eta^{\overline{\imath}}$, $\theta_i$ have nontrivial
boundary conditions in the open string.
Specifically, $\eta^{\overline{\imath}} = 0$
for Dirichlet directions,
and $\theta_i = \left( \mbox{Tr } F_{i \overline{\jmath}} \right) \eta^{
\overline{\jmath}}$ for Neumann conditions \cite{abooetal}.
As in \cite{orig}, we find it useful to temporarily make the
approximation that $\theta_i = 0$ for Neumann conditions.

As in \cite{dougmoore}, the orbifold spectrum consists of the
$G$-invariant states of this type.  Thus, as in \cite{orig}, to a
first approximation, we have described the massless boundary Ramond
sector states as $G$-invariant\footnote{There is an essentially
irrelevant technical subtlety we are glossing over here, involving the
$G$-action on the Fock vacua \cite{dixonthes}.  If a free complex
fermion transforms under $g \in G$ as $\psi \mapsto \exp(i \alpha)
\psi$, and if we label the Fock vacua corresponding to that single
fermion as $|\pm\rangle$, then naively under $g$, $|\pm\rangle \mapsto
\exp(\pm i \alpha / 2) | \pm \rangle$, which would be inconsistent
with our analysis above as the state corresponding to our $|0\rangle$
would no longer necessarily be $G$-invariant.  We are rescued by a
subtlety involving the square-root branch cut, which after work
described in \cite{dixonthes}, implies that our $|0\rangle$ above
actually is $G$-invariant.  We shall not consider this subtlety any
further in this paper, and will simply assume without further
discussion that any $|0\rangle$ which is not some meromorphic section
of some line bundle is always $G$-invariant.} parts of the sheaf
cohomology groups
\begin{displaymath}
H^n\left(S, {\cal E}^{\vee} \otimes {\cal F} \otimes \Lambda^m {\cal N}_{S/X}
\right)^G
\end{displaymath}
which are equivalent to the sheaf cohomology groups
\begin{displaymath}
H^n\left( [S/G], {\cal E}^{\vee} \otimes {\cal F} \otimes \Lambda^m {\cal N}_{S/X}
\right)
\end{displaymath}
on the quotient stack $[S/G]$,
since differential forms on $[X/G]$ are equivalent to $G$-invariant
differential forms on $X$.
(The group action on the space $X$ induces a canonical $G$-equivariant
structure on the normal bundle ${\cal N}_{S/X}$.)

As the reader might expect from \cite{orig}, there
is a spectral sequence
\begin{displaymath}
H^n\left(S, {\cal E}^{\vee} \otimes {\cal F} \otimes \Lambda^m {\cal N}_{S/X}
\right)^G \: \Longrightarrow \:
\mbox{Ext}^{n+m}_{ [X/G] }\left( i_* {\cal E}, i_* {\cal F} \right)
\end{displaymath}
(See appendix~\ref{pfs} for a proof.)
In passing, note this spectral sequence can be equivalently written as
\begin{displaymath}
H^n\left( [S/G], {\cal E}^{\vee} \otimes {\cal F} \otimes \Lambda^m {\cal N}_{S/X}
\right) \: \Longrightarrow \:
\mbox{Ext}^{n+m}_{ [X/G] }\left( i_* {\cal E}, i_* {\cal F} \right)
\end{displaymath}
to emphasize the relationship with stacks.

How is this spectral sequence realized physically?
Precisely as in \cite{orig}, there are two important subtleties
that must be taken into account:
\begin{enumerate}
\item $TX|_S$ need not split holomorphically, {\it i.e.} $TX|_S
\not\cong TS \oplus {\cal N}_{S/X}$ in general, so it is not quite right
to interpret the vertex operators as ${\cal N}_{S/X}$-valued forms.
\item The boundary conditions are twisted by the curvature of the
Chan-Paton factors as in \cite{abooetal}, so that for Neumann directions,
$\theta_i \neq 0$, but rather $\theta_i = \left( \mbox{Tr } F_{i \overline{\jmath}} \right) 
\eta^{ \overline{\jmath} }$.
\end{enumerate}
Just as in \cite{orig}, in the special case that ${\cal E} = {\cal F}$,
it can be seen explicitly  
that taking care of these two subtleties has the
effect of physically realizing the spectral sequence in BRST cohomology.

The physical realization of the spectral sequence is essentially
identical to that discussed in \cite{orig}.
As the method is essentially identical to that in
\cite{orig}, we shall only outline it here; see instead \cite{orig} for further
details.
Consider, for example, the differential 
\begin{displaymath}
d_2: \: H^0\left( S, {\cal E}^{\vee} \otimes {\cal E} \otimes
{\cal N}_{S/X} \right)^G
\: \longrightarrow \: H^2\left(S, {\cal E}^{\vee} \otimes {\cal E} \right)^G
\end{displaymath}
This differential turns out to be the composition
\begin{displaymath}
 H^0\left( S, {\cal E}^{\vee} \otimes {\cal E} \otimes
{\cal N}_{S/X} \right)^G
\: \stackrel{\alpha}{\longrightarrow} \:
H^1\left( S, {\cal E}^{\vee} \otimes {\cal E} \otimes TS \right)^G
\: \stackrel{\beta}{\longrightarrow} \:
 H^2\left(S, {\cal E}^{\vee} \otimes {\cal E} \right)^G
\end{displaymath}
(almost identical to the result for $d_2$ appearing in \cite{orig}).
The first map, $\alpha$, in the composition is the coboundary map in
the long exact sequence associated to the extension
\begin{displaymath}
0 \: \longrightarrow \: TS \: \longrightarrow \:
TX|_S \: \longrightarrow \: {\cal N}_{S/X} \:
\longrightarrow \: 0
\end{displaymath}
The second map, $\beta$, is the evaluation map\footnote{As a technical aside,
this map can
be described purely cohomologically. It is the cup product with the
Atiyah class $a({\mathcal E}) \in H^{1}(S,T^*S\otimes
End({\mathcal E},{\mathcal
E}))$ of ${\mathcal E}$. 
In particular, it shows that the second map has the same geometric
structure as the first one. The first map is the cup product with the
obstruction to linearizing the formal neighborhood of $S$ in $X$ and
the second map is the obstruction to liftability of the infinitesimal
symmetries of $S$ to infinitesimal symmetries of ${\mathcal
E}$. }, that takes a $TS$-valued index $\theta_i$ and replaces it
with a one-form times the trace of the curvature of ${\cal E}$, {\it
i.e.}, $\left( \mbox{Tr } F_{i \overline{\jmath}} \right)
\eta^{\overline{\jmath}}$.

Physically, since the correct boundary condition has $\theta_i \neq 0$
in general for Neumann directions, the forms in equation~(\ref{ramond1})
are valued in merely $\Lambda^m TX|_S$, instead of $\Lambda^m {\cal N}_{S/X}$.
In the present example, given an element of $H^0(S, {\cal E}^{\vee}
\otimes {\cal E} \otimes {\cal N}_{S/X} )$,
we can lift the coefficients from ${\cal N}_{S/X}$ to $TX|_S$,
and the result can be identified with a Ramond sector state of
the form~(\ref{ramond1}).
The BRST cohomology operator acts as $\overline{\partial}$, as usual.
The BRST operator does not quite annihilate our new $TX|_S$-valued forms,
but rather, produces $TS$-valued forms, physically
realizing the coboundary map $\alpha$ that forms the first part
of the differential $d_2$.
Now, the fact that the BRST operator does not quite
annihilate our states, but rather produces $TS$-valued forms,
would be a problem
were it not for the boundary condition on the $\theta_i$.
Applying the boundary condition on the $\theta_i$ is equivalent
to applying the evaluation map, which forms the second map $\beta$ inside
$d_2$.
Thus, we see that the differential $d_2$, described as the
composition of a coboundary map and an evaluation map,
is realized directly in BRST cohomology.

Proceeding in this fashion, just as in \cite{orig}, we see
explicitly that when ${\cal E} = {\cal F}$, the spectral
sequence relating normal-bundle-valued sheaf cohomology
to Ext groups on quotient stacks is realized physically
in terms of BRST cohomology.

Furthermore, we conjecture that the same result is true for all other cases,
not just ${\cal E} = {\cal F}$:
being careful about these subtleties should have the effect of realizing
all spectral sequences we discuss in this paper directly in BRST cohomology,
so that the physical massless boundary Ramond sector states are always in 
one-to-one correspondence with
elements of Ext groups on quotient stacks.

It is important to note that we are {\it not} assuming that string
orbifolds are the same as strings propagating on quotient stacks.
Rather, we have been doing standard first-principles computations directly
in BCFT, and have
discovered at the end of the day that the spectrum of massless 
boundary Ramond sector states
is counted by Ext groups on quotient stacks.  Nowhere do we make
any assumptions about physical relevance of stacks.
Of
course, this result is certainly evidence for the proposal that string
orbifolds describe strings on quotient stacks.

\subsection{Fractional branes, or, the necessity of stacks}

Let us take a moment to consider fractional branes.
So far in this paper we have used quotient stacks merely as
formal mathematical tools for manipulating orbifolds.
This might appear to the reader to be unnecessary -- surely,
one might claim,
this could all be done on quotient spaces, omitting the
technical complications of stacks.
However, in the process of describing fractional branes,
we shall see that if one wishes to describe branes in terms
of sheaves, then in order to describe fractional branes using
sheaves, one is forced to necessarily use stacks, as
many fractional branes cannot be described as sheaves
on quotient spaces.

Let us consider, for example, the simplest nontrivial fractional brane,
in the standard ALE orbifold of ${\bf C}^2$ by ${\bf Z}_2$.
The nontrivial fractional brane is a D0-brane on
the covering space ${\bf C}^2$ at the fixed point
of the ${\bf Z}_2$ action.  The Chan-Paton factors for the
nontrivial fractional brane have a nontrivial
orbifold group action, defined by the nontrivial irreducible representation
of ${\bf Z}_2$.  

In terms of sheaves, this D0-brane is described by a skyscraper sheaf
at the origin of ${\bf C}^2$.  This sheaf has two ${\bf
Z}_2$-equivariant structures, one for each irreducible representation
of ${\bf Z}_2$ acting on the generator of the module.  The equivariant
structure corresponding to the nontrivial fractional brane is defined
by the irreducible representation that multiplies the generator of the
module by negative one.  More formally, if we write
\begin{displaymath}
{\bf C}^2 \: = \: \mbox{Spec } {\bf C}[x,y]
\end{displaymath}
then the skyscraper sheaf is defined by the ${\bf C}[x,y]$-module
\begin{displaymath}
M \: \equiv \: {\bf C}[x,y]/(x,y) \: = \: {\bf C}[x,y] \cdot \alpha
\end{displaymath}
where $\alpha$ is the generator of the module,
and $x \cdot \alpha = y \cdot \alpha = 0$.
The possible equivariant structures on this sheaf are defined
by the possible ${\bf Z}_2$-actions on $\alpha$.
The two possible actions are obviously $\alpha \mapsto + \alpha$
and $\alpha \mapsto - \alpha$.

Now, let us project such sheaves to the quotient space.
For the trivial equivariant structure, defined by
$\alpha \mapsto + \alpha$, the module of ${\bf Z}_2$-invariants is given by
\begin{displaymath}
M^{ {\bf Z}_2 } \: = \: {\bf C}[x^2, y^2, xy] \cdot \alpha \: = \:
{\bf C} \cdot \alpha
\end{displaymath}
which defines a skyscraper sheaf at the origin of ${\bf C}^2 / {\bf Z}_2
= \mbox{Spec } {\bf C}[x^2,y^2,xy]$.
For the nontrivial equivariant structure, 
defined by
$\alpha \mapsto - \alpha$, corresponding to our nontrivial
fractional brane of interest, the module of ${\bf Z}_2$-invariants is
given by
\begin{displaymath}
M^{ {\bf Z}_2 } \: = \: 0.
\end{displaymath}
This module necessarily vanishes, because any ${\bf Z}_2$-invariants
are generated by $x \cdot \alpha$ and $y \cdot \alpha$, yet
both of these vanish.

Thus, we see that when we go all the way down to the quotient space,
the ${\bf Z}_2$-equivariant sheaf corresponding to the fractional
brane disappears.
More generally, working with orbifolds ({\it i.e.}, working on covering
spaces and manipulating invariant and equivariant objects)
is {\it not} equivalent to working on the quotient space.
In particular, we see here that the map from the orbifold
construction ({\it i.e.}, the quotient stack) to the quotient space
loses information.

In other words, in order to describe fractional branes using sheaves,
one is necessarily forced to work on a quotient stack,
and not on a quotient space.

\subsection{Examples}

In this section we shall discuss open string orbifold spectra
in a few examples.

\subsubsection{ $[ {\bf C}^3 / {\bf Z}_3 ]$ }  \label{c3z3}

To begin, consider a set of D0 branes at the origin of ${\bf C}^3$, and
consider the ${\bf Z}_3$ quotient of ${\bf C}^3$ that leaves
a single fixed point at the origin.
Let $\rho_0$, $\rho_1$, $\rho_2$ denote the three (one-dimensional)
irreducible representations of ${\bf Z}_3$, with $\rho_0$ the trivial
representation.

In order to talk about Ext groups between brane configurations,
we need to define two sets of branes.
The first set
will consist of $a+b+c$ D0 branes, all coincident
at the origin of ${\bf C}^3$.  In order to construct the orbifold,
we need to specify an equivariant structure on these branes,
{\it i.e.}, a group action on the Chan-Paton factors,
which in this simple case means we need to choose an $(a+b+c)$-dimensional
representation of ${\bf Z}_3$.  Put them in the $\rho_0^{\oplus a}
\oplus \rho_1^{\oplus b} \oplus \rho_2^{\oplus c}$ representation.
Let ${\cal E} = {\cal O}^{\oplus a+b+c}$ denote the (trivial) sheaf
over the origin of ${\bf C}^3$, and $i_* {\cal E}$ the corresponding
coherent sheaf on ${\bf C}^3$ (or $[ {\bf C}^3/{\bf Z}_3]$, depending
upon context).

Define a second set of D0 branes in a  similar way.
Put $a'+b'+c'$ D0 branes at the origin of ${\bf C}^3$, and define
a ${\bf Z}_3$-equivariant structure on the corresponding coherent
sheaf by picking the representation $\rho_0^{\oplus a'} \oplus
\rho_1^{\oplus b'} \oplus \rho_2^{\oplus c'}$ of ${\bf Z}_3$.
Let ${\cal F} = {\cal O}^{\oplus a'+b'+c'}$ denote the
corresponding sheaf over the origin of ${\bf C}^3$, and
$i_* {\cal F}$ the corresponding coherent sheaf on ${\bf C}^3$
or $[{\bf C}^3/{\bf Z}_3]$.

The quotient stack Ext's are straightforward to compute.
The normal bundle ${\cal N}_{pt/{\bf C}^3} = {\cal O}^3$,
and has\footnote{The ${\bf Z}_3$-equivariant structure on the normal
bundle is induced by the ${\bf Z}_3$ action on the ambient space;
it is not arbitrary.} ${\bf Z}_3$-equivariant structure given by
$\rho_1^{\oplus 3}$.
The bundle ${\cal E}^{\vee} \otimes {\cal F} = {\cal O}^{
(a+b+c)(a'+b'+c')}$, with ${\bf Z}_3$-equivariant structure given by
\begin{displaymath}
\rho_0^{\oplus a a' + b b' + c c'} \oplus
\rho_1^{\oplus a b' + c a' + b c'} \oplus
\rho_2^{\oplus b a' + a c' + c b'}
\end{displaymath}
(Using the fact that the dual of the trivial bundle ${\cal O}$ with
${\bf Z}_3$-equivariant structure defined by $\rho_i$
is the trivial bundle with ${\bf Z}_3$-equivariant structure
$\rho_i^{\vee} = \rho_{3-i}$, with indices taken mod 3.)
It is now straightforward to compute
\begin{eqnarray*}
H^0\left(pt, {\cal E}^{\vee} \otimes {\cal F} \right)^{{\bf Z}_3}
& = & {\bf C}^{a a' + b b' + c c'}, \\
H^0\left(pt, {\cal E}^{\vee} \otimes {\cal F} \otimes {\cal N}_{pt/{\bf C}^3}
\right)^{{\bf Z}_3} & = &
{\bf C}^{ 3 b a' + 3 a c' + 3 c b' }, \\
H^0\left(pt, {\cal E}^{\vee} \otimes {\cal F} \otimes \Lambda^2 {\cal N}_{pt/
{\bf C}^3} \right)^{{\bf Z}_3} & = &
{\bf C}^{ 3 a b' + 3 c a' + 3 b c'  }, \\
H^0\left(pt, {\cal E}^{\vee} \otimes {\cal F} \otimes \Lambda^3 {\cal N}_{pt/
{\bf C}^3} \right)^{{\bf Z}_3} & = &
{\bf C}^{a a' + b b' + c c'}.
\end{eqnarray*} 
Because the sheaf cohomology groups are computed over a point,
the spectral sequence degenerates, and we immediately derive
\begin{eqnarray*}
\mbox{Ext}^0_{ [ {\bf C}^3/{\bf Z}_3 ] } \left(
i_* {\cal E}, i_* {\cal F} \right) & = &
{\bf C}^{a a' + b b' + c c'}, \\
\mbox{Ext}^1_{ [ {\bf C}^3 / {\bf Z}_3 ] } \left(
i_* {\cal E}, i_* {\cal F} \right) & = &
{\bf C}^{  3 b a' + 3 a c' + 3 c b' }, \\
\mbox{Ext}^2_{ [ {\bf C}^3 / {\bf Z}_3 ] } \left(
i_* {\cal E}, i_* {\cal F} \right) & = &
{\bf C}^{ 3 a b' + 3 c a' + 3 b c'  }, \\
\mbox{Ext}^3_{ [ {\bf C}^3 / {\bf Z}_3 ] } \left(
i_* {\cal E}, i_* {\cal F} \right) & = &
{\bf C}^{a a' + b b' + c c'}. 
\end{eqnarray*}

We can compare these Ext groups against the known matter spectrum
in such ${\bf Z}_3$ orbifolds.  The Ext groups of degree zero are
correctly counting gauginos, and the Ext groups of degree one are
correctly counting the Higgsinos.
(Recall from \cite{orig}, however, that in general the type of
low-energy matter content is not determined by the degree of the
Ext group.  We shall see an example in this paper where
such relations break down when studying the ADHM construction
on ALE spaces.)

We also claimed earlier that in general,
\begin{displaymath}
\mbox{Ext}^i_{ [X/G] } \left( {\cal S}, {\cal T} \right) \: = \:
\mbox{Ext}^i_X \left( {\cal S}, {\cal T} \right)^G
\end{displaymath}
for any two $G$-equivariant sheaves ${\cal S}$, ${\cal T}$ on $X$.
That statement is easy to check in this case.
For example, if we compute Ext groups on $X$, using the methods
in \cite{orig}, we find
\begin{displaymath}
\mbox{Ext}^0_X\left( i_* {\cal E}, i_* {\cal F} \right) \: = \:
{\bf C}^{a a' + b b' + c c'} \oplus
{\bf C}^{a b' + c a' + b c' } \oplus
{\bf C}^{b a' + a c' + c b'}
\end{displaymath}
{}From the ${\bf Z}_3$-equivariant structure of
${\cal E}^{\vee} \otimes {\cal F}$, it is easy to see that
the first ${\bf C}$ factor is invariant under the ${\bf Z}_3$;
the second transforms as $\rho_1$; the third transforms as $\rho_2$.
The ${\bf Z}_3$ invariants are therefore given by
${\bf C}^{a a' + b b' + c c'}$, so in this example we see explicitly that
\begin{displaymath}
\mbox{Ext}^0_{ [ {\bf C}^3 / {\bf Z}_3 ]}
\left( i_* {\cal E}, i_* {\cal F} \right) \: = \:
\mbox{Ext}^0_{ {\bf C}^3 }
\left( i_* {\cal E}, i_* {\cal F} \right)^{ {\bf Z}_3 },
\end{displaymath}
a special case of the general relation above.

\subsubsection{ $[ {\bf C}^2 / {\bf Z}_n ]$ }  \label{c2zn}

Next, consider two sets of D0 branes at the fixed point of the ${\bf
Z}_n$ action on ${\bf C}^2$.  As described above, in order to make
sense of an orbifold of this theory, we need to specify a ${\bf
Z}_n$-equivariant structure on the D0 branes at the origin ({\it
i.e.}, the group action on the Chan-Paton factors).  Let $\rho_i$
label the $i$th one-dimensional irreducible representation of ${\bf
Z}_n$, so that $\rho_i(j)$ acts as multiplication by $\zeta^{ij}$,
where $\zeta={\rm exp}(2\pi i/n)$.  In particular, $\rho_0$ is the
trivial representation.  Fix the equivariant structure\footnote{ Since
we are dealing with equivariant sheaves over a point here, the
equivariant structure is completely determined by a choice of
representation of ${\bf Z}_n$.}  on the first set by taking $a_i$ D0
branes to be in the $\rho_i$ representation, and fix the equivariant
structure on the second set by taking $b_i$ D0 branes to be in the
$\rho_i$ representation.  Let ${\cal E}$ and ${\cal F}$ denote the
corresponding sheaves over the origin of ${\bf C}^2$, so that ${\cal
E} = {\cal O}^{\oplus a_0 + a_1 + \cdots + a_{n-1}}$ and similarly for
${\cal F}$.  The ${\bf Z}_n$ action is understood but is suppressed
from this notation.  Using the fact that the induced equivariant
structure on the normal bundle ${\cal O} \oplus {\cal O}$ is given by
$\rho_1 \oplus \rho_{n-1}$, and that $\rho_i^{\vee} = \rho_{n-i}$, it
is straightforward to compute
\begin{eqnarray*}
H^0\left(pt, {\cal E}^{\vee} \otimes {\cal F} \right)^{{\bf Z}_n}
& = & {\bf C}^{a_0 b_0 + a_1 b_1 + a_2 b_2 + \cdots + a_{n-1} b_{n-1} }, \\
H^0\left(pt, {\cal E}^{\vee} \otimes {\cal F} \otimes {\cal N}_{pt/{\bf C}^2} 
\right)^{ {\bf Z}_n}
& = & {\bf C}^{a_0 b_{n-1} + a_1 b_0 + a_2 b_1 + \cdots + a_{n-2} b_{n-3} +  a_{n-1} b_{n-2}} \\
& & \oplus {\bf C}^{a_{n-1} b_0 + a_0 b_1 + a_1 b_2 + \cdots + a_{n-2} b_{n-1} }, \\
H^0\left(pt, {\cal E}^{\vee} \otimes {\cal F} \otimes \Lambda^2 {\cal N}_{
pt/{\bf C}^2} \right)^{ {\bf Z}_n}
& = & H^0\left(pt, {\cal E}^{\vee} \otimes {\cal F} \right)^{{\bf Z}_n}, \\
& = &  {\bf C}^{a_0 b_0 + a_1 b_1 + a_2 b_2 + \cdots + a_{n-1} b_{n-1}}, 
\end{eqnarray*}
(using the fact that $\Lambda^2 {\cal N}_{ pt/{\bf C}^2} = {\cal O}$ with 
equivariant structure defined by $\rho_1 \otimes \rho_{n-1} = \rho_0$).
Since $S$ is a point, the spectral sequence degenerates,
so we can read off that
\begin{eqnarray*}
\mbox{Ext}^0_{ [ {\bf C}^2 / {\bf Z}_n ] } \left( i_* {\cal E}, i_* {\cal F}
\right) & = & {\bf C}^{a_0 b_0 + a_1 b_1 + a_2 b_2 + \cdots + a_{n-1} b_{n-1}}, \\
\mbox{Ext}^1_{  [ {\bf C}^2 / {\bf Z}_n ] } \left( i_* {\cal E}, i_* {\cal F}
\right) & = &
{\bf C}^{a_0 b_{n-1} + a_1 b_0 + a_2 b_1 + \cdots + a_{n-2} b_{n-3} +  a_{n-1} b_{n-2}} \\
& & \oplus {\bf C}^{a_{n-1} b_0 + a_0 b_1 + a_1 b_2 + \cdots + a_{n-2} b_{n-1} }, \\
\mbox{Ext}^2_{  [ {\bf C}^2 / {\bf Z}_n ] } \left( i_* {\cal E}, i_* {\cal F}
\right) & = &
{\bf C}^{a_0 b_0 + a_1 b_1 + a_2 b_2 + \cdots + a_{n-1} b_{n-1} }. 
\end{eqnarray*}

\subsubsection{ $[ {\bf C}^2 / {\bf Z}_n ]$ revisited}

Next, we shall consider again D0 branes on $[ {\bf C}^2 / {\bf Z}_n ]$,
but this time, we shall displace the D0 branes away from the singularity.

Consider $n$ D0 branes on ${\bf C}^2$, 
located at the $n$ points $(x_i, y_i) = (\xi^i, \xi^{-i})$,
where $\xi = \exp(2 \pi i / n)$.
Under the ${\bf Z}_n$ action $(x,y) \mapsto (\xi x, \xi^{n-1} y)$,
the D0 branes are permuted among themselves.
In order to define the orbifold, we pick the natural
group action on the Chan-Paton factors
({\it i.e.} a ${\bf Z}_n$-equivariant structure on the
corresponding coherent sheaf) induced by the natural action of ${\bf Z}_n$
on ${\bf C}^2$.

Define $S$ to be the disjoint union of the
points $(x_i, y_i) = (\xi^i, \xi^{-i})$ on ${\bf C}^2$. 
In the language we have been using, we are considering D0 branes
wrapped on $S$.
Note that the ${\bf Z}_n$ action on ${\bf C}^2$ maps $S$ back into
itself.

Let us compute the massless boundary Ramond sector spectrum
in open strings stretching from these D0 branes back to themselves.
{}From our general arguments, that spectrum is given by
$\mbox{Ext}^*_{ [ {\bf C}^2 / {\bf Z}_n ] } \left( {\cal O}_S,
{\cal O}_S \right)$.

Proceeding as usual, we first need to calculate the ${\bf Z}_n$-invariants
of the sheaf cohomology groups
\begin{displaymath}
H^0\left(S, {\cal O}_S^{\vee} \otimes {\cal O}_S \otimes
\Lambda^p {\cal N}_{S/{\bf C}^2} \right) \: = \:
\left\{ \begin{array}{cl}
        {\bf C}^n & p = 0,2, \\
        {\bf C}^{2n} & p=1.
        \end{array}  \right.
\end{displaymath}
In order to work out the ${\bf Z}_n$-invariants, we need to
understand the induced ${\bf Z}_n$-equivariant structure on
$N_{S/{\bf C}^2} = {\cal O}_S \oplus {\cal O}_S$.
        
Now, the generator of the ${\bf Z}_n$ action maps 
\begin{displaymath}
\left(\, x,\: y\, \right) \: \mapsto \: \left( \,
\xi x, \: \xi^{-1} y \, \right).
\end{displaymath}
If we translate the origin to lie on one of the points in $S$,
we see that the generator of ${\bf Z}_n$ maps
\begin{displaymath}
\left( \, x \: - \: \xi^i , \: y \: - \: \xi^{-i} \, \right)
\: \mapsto \:
\left( \, \xi x \: - \: \xi^{i+1}, \:
\xi^{-1} y \: - \: \xi^{-i-1} \, \right)
\end{displaymath}
In other words, the generator of the ${\bf Z}_n$ action simultaneously
acts by phases on the sections of the normal bundle while exchanging points.
In terms of the data above, the sections of 
${\cal O}_S^{\vee} \otimes {\cal O}_S$ 
and ${\cal O}_S^{\vee} \otimes {\cal O}_S \otimes \Lambda^2 {\cal N}_{S / {\bf C}^2}$
are cyclically permuted
by the action of ${\bf Z}_n$,
{\it i.e.} the generator of ${\bf Z}_n$ acts on a set of basis
elements of the vector space of sections as
\begin{displaymath}
\left( s_1, \cdots s_n \right) \: \mapsto \:
\left( s_n, s_1, s_2, \cdots, s_{n-1} \right)
\end{displaymath}
The sections of ${\cal O}_S^{\vee} \otimes {\cal O}_S \otimes {\cal N}_{S/
{\bf C}^2 }$ are cyclically permuted and also multiplied by $(\xi, \xi^{-1})$.
In more detail, these sections are $n$-tuples of pairs of complex numbers.
Each pair is the section over one connected component of $S$
({\it i.e.}, over a single point), and there are $n$ components (points),
hence we have an $n$-tuple of data.
The generator of ${\bf Z}_n$ maps such an $n$-tuple 
\begin{displaymath}
\left( \, (s_1, t_1), \: (s_2, t_2), \: \cdots, \: (s_n, t_n) \, \right)
\end{displaymath}
to the $n$-tuple given by
\begin{displaymath}
\left( \, (\xi s_n, \xi^{-1} t_n), \: (\xi s_1, \xi^{-1} t_1), \:
(\xi s_2, \xi^{-1} t_2), \: \cdots, \: (\xi s_{n-1}, \xi^{-1} t_{n-1} )
\, \right).
\end{displaymath}

Next, we need to calculate the ${\bf Z}_n$ invariants.
The only way to construct a ${\bf Z}_n$ invariant from a set of
objects that are cyclically permuted by ${\bf Z}_n$ is to take
their sum.  In other words, using the terminology above,
the ${\bf Z}_n$ invariant sections of ${\cal O}_S^{\vee} \otimes
{\cal O}_S$ have the single generator $(s_1 + s_2 + \cdots + s_n)$.
With a little thought, it is straightforward to see that the invariant
sections of ${\cal O}_S^{\vee} \otimes {\cal O}_S \otimes {\cal N}_{S / {\bf C}^2}$
are generated by the following two $n$-tuples of pairs of
complex numbers:
\begin{displaymath}
\begin{array}{c}
\left( \, (1,0), \: (\xi,0), \: (\xi^2,0), \: \cdots, (\xi^{n-1},0) \, \right),
 \\
\left( \, (0,1), \: (0, \xi^{-1}), \: (0, \xi^{-2}), \: 
\cdots, \: (0, \xi^{-n-1}) \, \right)
\end{array}
\end{displaymath}

Thus, the ${\bf Z}_n$-invariants of the
sheaf cohomology groups above are given by
\begin{displaymath}
H^0\left(S, {\cal O}_S^{\vee} \otimes {\cal O}_S \otimes
\Lambda^p {\cal N}_{S/{\bf C}^2} \right)^{ {\bf Z}_n } \: = \:
\left\{ \begin{array}{cl}
        {\bf C} & p=0, 2, \\
        {\bf C}^2 & p=1, \\
        0 & \mbox{otherwise}. 
        \end{array}  \right.
\end{displaymath}
Since the sheaf cohomology groups are computed over points,
the spectral sequence degenerates, and so we have that
\begin{displaymath}
\mbox{Ext}^p_{ [ {\bf C}^2 / {\bf Z}_n ] } \left( {\cal O}_S,
{\cal O}_S \right) \: = \:
\left\{ \begin{array}{cl}
        {\bf C} & p=0, 2, \\
        {\bf C}^2 & p=1, \\
        0 & \mbox{otherwise}. 
        \end{array}  \right.
\end{displaymath}

Now, let us check our result against what we would expect
physically.
This set of $n$ D0 branes on ${\bf
C}^2$, away from the singularity, descends to a single D0 brane on the
quotient located away from the singularity.  So, we should be
calculating the spectrum of open string states between a D0 brane and
itself, on a smooth patch.  This spectrum we already know, from for
example \cite{orig} -- we should find Ext$^0$ and Ext$^2$ have
dimension one, and Ext$^1$ has dimension two, precisely as we observe
above.  In other words, we should get one six-dimensional vector
multiplet, corresponding to the $U(1)$ gauge symmetry of the brane,
and containing a single six-dimensional spinor, and one
six-dimensional hypermultiplet, also containing a single\footnote{In
four dimensions, hypermultiplets contain two spinors, but in six
dimensions they contain only one.  Note that this is a necessary
consequence of the consistency of dimensional reduction.}
six-dimensional spinor.  Since Serre duality maps the Ext groups above
back into themselves, with only a degree shift, the Ext groups above
should contain the spectra for both orientations.  In short, the Ext
groups we have calculated above are precisely right to agree with what
we expect on physical grounds.

\subsubsection{An example with a nonabelian orbifold group }

So far all of our examples have involved abelian orbifold groups.
Just to make the point that our analysis applies to both
abelian and nonabelian orbifold groups, we shall now consider
two-dimensional $D_n$ orbifolds.

A two-dimensional $D_{n+2}$ orbifold is defined by the
binary dihedral subgroup $BD_{4n}$ of $SU(2)$.
The action of this finite group on ${\bf C}^2$
is generated by the two matrices \cite{reid1}
\begin{displaymath}
\alpha \: = \: \left( \begin{array}{cc}
                      \xi & 0 \\
                      0 & \xi^{-1}
                      \end{array}  \right), \:
\beta \: = \: \left( \begin{array}{cc}
                     0 & 1 \\
                     -1 & 0
                     \end{array} \right)
\end{displaymath}
where $\xi = \exp( 2\pi i / 2n)$.

The finite group $BD_{4n}$ has $n-1$ two-dimensional irreducible
representations, and four one-dimensional irreducible representations,
which are conveniently expressed in terms of a set of $n+1$ 
(not necessarily irreducible) two-dimensional representations.
Let $\rho_i$ denote the $i$th
such two-dimensional representation, for $0 \leq i \leq n$.
The generators $\alpha$ and $\beta$
are expressed in the representation $\rho_i$ by \cite{reid1}
\begin{displaymath}
\alpha \: = \: \left( \begin{array}{cc}
                      \xi^i & 0 \\
                      0 & \xi^{-i} 
                      \end{array}  \right), \:
\beta \: = \: \left( \begin{array}{cc}
                     0 & 1 \\
                     (-1)^i & 0 
                     \end{array}  \right)
\end{displaymath}
The representations $\rho_i$ for $1 \leq i \leq n-1$ are irreducible
two-dimensional representations; the representations $\rho_0$ and $\rho_n$
each decompose into a pair of one-dimensional representations.
For $0 < i < n$, the representations obey $\rho_i \otimes \rho_1 = 
\rho_{i-1} \oplus \rho_{i+1}$.

Let us suppose we have two sets of D0 branes, located
at the origin of ${\bf C}^2$.
The first set will consist of $a$ D0 branes, in the
trivial one-dimensional representation of $BD_{4n}$.
Denote the corresponding trivial rank $a$ bundle over the
origin of ${\bf C}^2$ by ${\cal E}$.
The second set will consist of $2b$ D0 branes, in the
$\rho_1^{\oplus b}$ representation of $BD_{4n}$,
and denote the corresponding trivial rank $2b$ bundle over the
origin of ${\bf C}^2$ by ${\cal F}$.

To calculate the open string spectrum, we now follow the prescription
given earlier. 
We first calculate sheaf cohomology on the covering space ${\bf C}^2$:
\begin{eqnarray*}
H^0\left(pt, {\cal E}^{\vee} \otimes {\cal F} \right) & = &
{\bf C}^{2ab}, \\
H^0\left( pt, {\cal E}^{\vee} \otimes {\cal F} \otimes {\cal N}_{pt/{\bf C}^2}
\right) & = &
{\bf C}^{4ab},
\end{eqnarray*}
where the first set of sections transforms as $\rho_1^{\oplus ab}$
under $BD_{4n}$, and the second set transforms as
$(\rho_1 \otimes \rho_1)^{\oplus ab}$,
using the fact that ${\cal N}_{pt/{\bf C}^2}$ has fiber ${\bf C}^2$,
and has $BD_{4n}$-equivariant structure defined by $\rho_1$.
Since the representation $\rho_1$ is irreducible and non-trivial,
there are no $BD_{4n}$-invariant sections of ${\cal E}^{\vee} \otimes
{\cal F}$.  In the second set, we use the fact that $\rho_1 \otimes \rho_1
= \rho_0 \oplus \rho_2$, and although $\rho_2$ is irreducible and
nontrivial, exactly one of the one-dimensional representations that
$\rho_0$ decomposes into is the trivial one-dimensional representation.
As a result,
\begin{eqnarray*}
H^0\left(pt, {\cal E}^{\vee} \otimes {\cal F} \right)^{ BD_{4n} } & = &
0, \\
H^0\left( pt, {\cal E}^{\vee} \otimes {\cal F} \otimes {\cal N}_{pt/{\bf C}^2}
\right)^{ BD_{4n} } & = &
{\bf C}^{ab}.
\end{eqnarray*}

Since we are computing sheaf cohomology over a point,
the spectral sequence degenerates, and so we can immediately read off that
\begin{eqnarray*}
\mbox{Ext}^0_{ [ {\bf C}^2 /  BD_{4n}  ]} \left(
i_* {\cal E}, i_* {\cal F} \right) & = & 0, \\
\mbox{Ext}^1_{ [ {\bf C}^2 / BD_{4n} ]} \left(
i_* {\cal E}, i_* {\cal F} \right) & = & {\bf C}^{ab}.
\end{eqnarray*}

\subsubsection{ A non-supersymmetric example: $ [ {\bf C} / {\bf Z}_n ]$ }
\label{czn}

Most of the examples we shall study will involve supersymmetric
orbifolds.  However, as we are merely calculating massless boundary Ramond
sector spectra, the same methods apply to nonsupersymmetric orbifolds
as well.  In particular, in this subsection we shall consider
the one-dimensional orbifold ${\bf C}/ {\bf Z}_n$, with the generator of
${\bf Z}_n$ acting as multiplication by ${\rm exp}(2\pi i/n)$.
 
One-dimensional quotients of this form are a little subtle,
in that not only does the group action not preserve a holomorphic
form, but also in that
they are essentially invisible in the algebraic geometry of
quotient spaces;
the coordinate ring for ${\bf C}/{\bf Z}_n$ is the same as
that for ${\bf C}$.
However, differential geometry certainly still sees the
singularity, and orbifolds of this form have recently
been popular in the literature \cite{evaetal,hkm,mm}.  
More to the point, these singularities are also very visible
in the algebraic geometry of quotient stacks.  We have here another
example, beyond fractional branes, where understanding the brane
construction in sheaf language necessarily involves stacks.

Technically, we are going to consider a very slightly different orbifold
than was considered in those papers (the difference involves the
group action on the fermions), so the spectra we will obtain will
be slightly different.

As in the previous examples in this section, we shall consider
open string spectra in orbifolds involving two sets of D0 branes,
both sitting at the fixed point of the group action on the covering
space.  Note that since we are in a non-supersymmetric situation,
we are counting the spectrum of massless spacetime fermions ({\it i.e.} massless
boundary Ramond sector states).  
As before, let $\rho_i$ denote irreducible representations
of ${\bf Z}_n$, with $\rho_0$ the trivial representation.

Let the first set of D0 branes be defined by $a_i$ D0 branes,
in the $\rho_i$ representation of ${\bf Z}_n$, for all $i$.
(This defines the group action on the Chan-Paton factors,
and equivalently, the $G$-equivariant structure on the corresponding
coherent sheaf.)
Let ${\cal E}$ denote the corresponding sheaf over the origin
of ${\bf C}$.

Let the second set of D0 branes be defined by $b_i$ D0 branes,
in the $\rho_i$ representation of ${\bf Z}_n$,
and let ${\cal F}$ denote the corresponding sheaf over the origin
of ${\bf C}$.

Using the fact that ${\cal N}_{pt/{\bf C}} = {\cal O}$, with ${\bf Z}_n$-equivariant
structure defined by $\rho_1$, and that $\rho_i^{\vee} = \rho_{n-i}$,
it is straightforward to calculate 
\begin{eqnarray*}
H^0\left(pt, {\cal E}^{\vee} \otimes {\cal F} \right)^{ {\bf Z}_n }
& = & {\bf C}^{a_0 b_0 + a_1 b_1 + \cdots + a_{n-1} b_{n-1} }, \\
H^0\left(pt, {\cal E}^{\vee} \otimes {\cal F} \otimes {\cal N}_{ pt/{\bf C}} 
\right)^{ {\bf Z}_n }
& = & {\bf C}^{a_1 b_0 + a_2 b_1 \cdots + a_{n-1} b_{n-2} + a_0 b_{n-1} }.
\end{eqnarray*}
Since these sheaf cohomology groups are computed over a point,
the spectral sequence degenerates, so
we derive
\begin{eqnarray}
\label{nonsusy}
\mbox{Ext}^0_{ [ {\bf C} / {\bf Z}_n ] } \left( i_* {\cal E},
i_* {\cal F} \right) & = &
{\bf C}^{a_0 b_0 + a_1 b_1 + \cdots + a_{n-1} b_{n-1} }, \\
\label{nonsusy2}
\mbox{Ext}^1_{ [ {\bf C} / {\bf Z}_n ] } \left( i_* {\cal E},
i_* {\cal F} \right) & = &
 {\bf C}^{a_1 b_0 + a_2 b_1 + \cdots + a_{n-1} b_{n-2} + a_0 b_{n-1} }. 
\end{eqnarray}

Using the obvious symmetry between ${\cal E}$ and ${\cal F}$, one
can also compute
\begin{eqnarray*}
\mbox{Ext}^0_{ [ {\bf C} / {\bf Z}_n ] } \left( i_* {\cal F},
i_* {\cal E} \right) & = &
{\bf C}^{a_0 b_0 + a_1 b_1 + \cdots + a_{n-1} b_{n-1} }, \\
& = & \mbox{Ext}^0_{ [ {\bf C}/ {\bf Z}_n ] } \left(
i_* {\cal E}, j_* {\cal F} \right), \\
\mbox{Ext}^1_{ [ {\bf C} / {\bf Z}_n ] } \left( i_* {\cal F},
i_* {\cal E} \right) & = &
{\bf C}^{a_0 b_1 + a_1 b_2 + \cdots + a_{n-2} b_{n-1} + a_{n-1} b_0 }.
\end{eqnarray*}

Let us check this result for consistency with Serre duality.
In supersymmetric one-dimensional
cases, Serre duality is the statement that
\begin{displaymath}
\mbox{Ext}^n \left( i_* {\cal E}, i_* {\cal F} \right) \: \cong \:
\mbox{Ext}^{1-n} \left( i_* {\cal F}, i_* {\cal E} \right)^\vee 
\end{displaymath}
which is certainly not what we see here.
The resolution of this difficulty lies in the fact that we
have a {\it non}-supersymmetric orbifold.
Although we started with a Calabi-Yau ({\it i.e.} ${\bf C}$),
we quotiented by a group action that does not preserve the
holomorphic top form.
As a result, the correct statement of Serre duality in this case is
that
\begin{equation}  \label{correctserre}
\mbox{Ext}^n_{ [ {\bf C} / {\bf Z}_n ] } \left( i_* {\cal E},
i_* {\cal F} \right) \: \cong \:
\mbox{Ext}^{1-n}_{ [ {\bf C} / {\bf Z}_n ] } \left( i_* {\cal F},
i_* {\cal E} \otimes K \right)^\vee
\end{equation}
where $K$ is the canonical bundle on ${\bf C}$.
Although the canonical bundle $K$ on ${\bf C}$ is trivial,
the ${\bf Z}_n$-equivariant structure induced on $K$
is {\it not} trivial in a non-supersymmetric orbifold
of a Calabi-Yau, and so $K$ cannot be omitted.

We can compute the Ext groups with $K$ directly, using our
spectral sequence, since $i_* {\cal E} \otimes K$ is
equivalent to tensoring ${\cal E}$ with a trivial rank 1 bundle
with nontrivial equivariant structure.  Since $K$ is dual
to the tangent bundle which corresponds to the representation
$\rho_1$, we see that $K$ corresponds to the representation 
$\rho_{n-1}=\rho_1^\vee$.
It is then straightforward to compute
\begin{eqnarray*}
\mbox{Ext}^0_{ [ {\bf C} / {\bf Z}_n ] } \left( i_* {\cal F},
i_* {\cal E} \otimes K \right) & = &
{\bf C}^{a_1 b_0 + a_2 b_1 + \cdots + a_{n-1} b_{n-2} + a_0 b_{n-1} }, \\
\mbox{Ext}^1_{ [ {\bf C} / {\bf Z}_n ] } \left( i_* {\cal F},
i_* {\cal E} \otimes K \right) & = &
{\bf C}^{a_0 b_0 + a_1 b_1 + \cdots + a_{n-1} b_{n-1} }, 
\end{eqnarray*}
in complete agreement with (\ref{nonsusy}), (\ref{nonsusy2}), and the duality
(\ref{correctserre}).

\subsubsection{Aside:  equivariant structures on nontrivial bundles}

So far we have only considered examples
in which all bundles appearing are trivial.
In the next few subsections we shall consider some less trivial
examples, involving curves in more general Calabi-Yau's,
so we shall take a moment to discuss some of the technical
complications that can pop up.

One important technical complication that can arise
involves $G$-equivariant structures, {\it i.e.}, the choice
of group action on the Chan-Paton factors.
So far this choice has simply amounted to a choice of representation
of the group, but that is only because all bundles we have
looked at so far were trivial.

In general, a $G$-equivariant structure on a bundle or sheaf ${\cal E}
\rightarrow M$
is a lift of the group action on the base, {\it i.e.}, for
all $g: M \rightarrow M$, a $G$-equivariant structure assigns
a $\tilde{g}: {\cal E} \rightarrow {\cal E}$ such that
the following diagram commutes:
\begin{displaymath}
\xymatrix{
{\cal E} \ar[r]^{ \tilde{g} } \ar[d] & {\cal E} \ar[d] \\
M \ar[r]^{g} & M 
}
\end{displaymath}
and such that, for any two group elements $g$, $h$ in $G$,
$\widetilde{gh} = \tilde{g} \circ \tilde{h}$.
In general, a $G$-equivariant structure on a given bundle or sheaf
need not exist; when they do exist, they are not unique.
(In the case of bundles, non-uniqueness is the statement that
any one $G$-equivariant structure can be combined with a suitable
set of gauge transformations to create another $G$-equivariant structure.)
In heterotic compactifications, the ambiguity in the choice of
$G$-equivariant structure is known historically as the choice
of orbifold Wilson lines.  The same ambiguity in the
$G$-equivariant structures for $B$ fields is known as discrete torsion
\cite{medt,dtrev}. 

If a bundle ${\cal E}$ is trivial, then $G$-equivariant structures
always exist, and are essentially determined by a representation of $G$.
If the bundle ${\cal E}$ is nontrivial, then $G$-equivariant structures
need not exist.  For a simple example, consider the nontrivial
principal ${\bf Z}_2$-bundle on $S^1$, {\it i.e.}, the boundary of
the M\"obius strip.
Let $G$ be the group of rotations of the circle.
In this case, there is no $G$-equivariant structure on the total space
of the bundle.  The problem is that one cannot satisfy the
condition $\widetilde{gh} = \tilde{g}\circ \tilde{h}$:  a rotation of the
$S^1$ by $360^{\circ}$ does not take you back to the place you
started on the M\"obius strip -- you have to rotate the $S^1$ twice
to return to your original position on the M\"obius strip.

In this paper, most of the examples we shall study will involve
bundles and sheaves that admit an equivariant structure, so for the most part
we shall ignore this issue in this paper.

\subsubsection{${\bf Z}_5$ action on the Fermat quintic threefold }

So far we have only considered topologically trivial examples,
involving quotients of affine spaces with isolated fixed points.
Next, we shall consider some topologically nontrivial examples.

Consider the Fermat quintic threefold, the hypersurface defined by
the equation
\begin{displaymath}
x_0^5 \: + \: x_1^5 \: + \: x_2^5 \: + \: x_3^5 \: + \: x_4^5 \: = \: 0
\end{displaymath}
in ${\bf P}^4$.
The Fermat quintic has families of rational curves.
One particular rational curve, call it $C$, is defined by the complete
intersection
\begin{displaymath}
\label{magnetic}
x_0 \: + \: x_1 \: = \:
x_2 \: + \: x_3 \: = \:
x_4 \: = \: 0
\end{displaymath}
and has normal bundle ${\cal N}_{C/X} =  {\cal O}(1) \oplus {\cal O}(-3)$.
For the convenience of the reader, we repeat the computation from 
\cite{finite}.  

If $L\subset X\subset {\bf P}^4$ 
is any line in any nonsingular quintic threefold with defining equation
$F(x_0,\ldots,x_4)=0$, the following computational scheme is given.
Identify $L$ with ${\bf P}^1$ via any parametrization.  Then consider the map
\[
\Phi_L:H^0({\cal O}_{{\bf P}^1}(1))^5\to H^0({\cal O}_{{\bf P}^1}(5))
\]
defined by 
\[
\Phi_L(\alpha_0,\ldots,\alpha_4)=\sum_{i=0}^4\alpha_i\frac{\partial F}
{\partial x_i}\mid_L.
\]
Infinitesimal reparametrizations of the ${\bf P}^1$ induce a ${\rm gl}(2)$
action on the kernel of $\Phi_L$.  The result of \cite{finite} is
\begin{eqnarray*}
H^0(L,{\cal N}_{L/X}) \simeq {\rm ker}(\Phi_L)/{\rm gl}(2),\\
H^1(L,{\cal N}_{L/X}) \simeq {\rm coker}(\Phi_L).
\end{eqnarray*}

In the case of the line given above, we can use $(x_1,x_3)$ as
homogeneous coordinates on $C\simeq{\bf P}^1$.  The map $\Phi_C$ can
be described by the matrix of restricted partial derivatives; up to an
overall factor of 5 this is
\[
\left(x_1^4,x_1^4,x_3^4,x_3^4,0\right).
\]
The cokernel of $\Phi_C$ 
is 2 dimensional, spanned by $\{x_1^3x_3^2,x_1^2x_3^3\}$.
We conclude that $H^1({\cal N}_{C/X})\simeq{\bf C}^2$.
Since ${\cal N}_{C/X}\simeq {\cal O}(a)\oplus {\cal O}(b)$ with $a+b=-2$, this
uniquely specifies ${\cal N}_{C/X} =  {\cal O}(1) \oplus {\cal O}(-3)$.

The kernel of $\Phi_C$ is 6 dimensional:
\[
{\rm ker}(\Phi_C)=\left\{\left(\ell_1,-\ell_1,\ell_2,-\ell_2,\ell_3\right)
\right\}
\]
where the $\ell_i$ are arbitrary homogeneous linear forms on ${\bf P}^1$.
The ${\rm gl}(2)$ action is transitive on the space of $\ell_1,\ell_2$,
so ${\rm ker}(\Phi_C)$ can be described by $\ell_3$.  From the computation
in \cite{finite}, the corresponding normal vector field is
\begin{equation}
\label{nvf}
v=\ell_3\frac{\partial}{\partial x_4}|_C.
\end{equation}

The Fermat quintic has a number of finite symmetry groups.
Consider, for example, the ${\bf Z}_5$ action on this quintic
described by the action of its generator on the homogeneous
coordinates of ${\bf P}^4$ as follows:
\begin{displaymath}
\left( x_0, x_1, x_2, x_3, x_4 \right) \: \mapsto \:
\left( x_0, x_1, \alpha x_2, \alpha x_3, \alpha^3 x_4 \right)
\end{displaymath}
where $\alpha$ is a fifth root of unity.
Under this ${\bf Z}_5$ action, the curve $C$ is mapped back into itself,
in the sense that although points on the curve may move around under the
${\bf Z}_5$, they will move to other points on the same curve;
the ${\bf Z}_5$ maps the curve back into itself (nontrivially).
The equivariant structure on the normal bundle ${\cal N}_{C/X}$ is such
that the two sections transform as $\alpha^2$ and
$\alpha^3$.  Indeed, from (\ref{nvf}) the sections are 
$x_1\partial/\partial x_4$ and $x_3\partial/\partial x_4$ which have the
indicated transformation properties.  Similarly, from the above discussion
we can also see that the two sections of $H^1({\cal N}_{C/X})$ transform
as $\alpha^2$ and $\alpha^3$ under ${\bf Z}_5$.

Now, wrap a single D2 brane on the fixed curve $C$, and assume
its gauge bundle is trivial.  Let us assume the group action
on the Chan-Paton factors ({\it i.e.} the ${\bf Z}_5$-equivariant
structure on the gauge bundle) is the one induced by the ${\bf Z}_5$
action on $X$, so in particular is trivial on global functions (which are
constants).
We shall now compute the massless
boundary Ramond sector spectrum of open strings connecting
that D2 brane to itself, in the ${\bf Z}_5$ orbifold above.
In other words, let us calculate $\mbox{Ext}^*_{ [ X / {\bf Z}_5 ] }
\left( {\cal O}_C, {\cal O}_C \right)$.
Following the methods we have introduced so far,
we first calculate some sheaf cohomology.
We shall first calculate the sheaf cohomology on the covering space,
and then describe the $G$-action on the sheaf cohomology groups.
Then, we shall describe the $G$-invariants, and finally 
describe the $\mbox{Ext}$ groups.

The first two relevant sheaf cohomology groups on the covering space 
are given by
\begin{eqnarray*}
H^0\left( C, {\cal O}_C^{\vee} \otimes {\cal O}_C \right)
& = & {\bf C}, \\
H^1\left(C, {\cal O}_C^{\vee} \otimes {\cal O}_C \right)
& = & 0.
\end{eqnarray*}
Since we picked the ${\bf Z}_5$-equivariant structure on ${\cal O}_C$
to be trivial, both of these sheaf cohomology groups are ${\bf Z}_5$-invariant.
The next two relevant sheaf cohomology groups are
\begin{eqnarray*} 
H^0\left(C, {\cal O}_C^{\vee} \otimes {\cal O}_C \otimes {\cal N}_{C/X} \right)
& = & H^0\left(C, {\cal O}_C(1) \right), \\
& = & {\bf C}^2, \\
H^1\left(C, {\cal O}_C^{\vee} \otimes {\cal O}_C \otimes {\cal N}_{C/X} \right)
& = & H^1\left(C, {\cal O}_C(1) \oplus {\cal O}_C(-3) \right), \\
& = & {\bf C}^2. 
\end{eqnarray*}
The first of these sheaf cohomology groups is given by $H^0(C, {\cal N}_{C/X})
= {\bf C}^2$,
which is not ${\bf Z}_5$-invariant.  Rather, the sections of
${\cal N}_{C/X}$ transform as $(\alpha^2, \alpha^3)$ under the ${\bf Z}_5$.
The second of these sheaf cohomology groups is given by $H^1(C, {\cal N}_{C/X})
 = {\bf C}^2$, and is also not ${\bf Z}_5$-invariant, and again the
sections transform as $(\alpha^2,\alpha^3)$.

The final two sheaf cohomology groups on the covering space
we shall require are given by
\begin{eqnarray*}
H^0\left(C, {\cal O}_C^{\vee} \otimes {\cal O}_C \otimes \Lambda^2 {\cal N}_{C/X}
\right) & = & H^0\left(C, {\cal O}_C(-2) \right), \\
& = & 0, \\
H^1\left(C, {\cal O}_C^{\vee} \otimes {\cal O}_C \otimes \Lambda^2 {\cal N}_{C/X}
\right) & = & H^1\left(C, {\cal O}_C(-2) \right), \\
& = & {\bf C}.
\end{eqnarray*}

The elements of $H^1(C, \Lambda^2 {\cal N}_{C/X})$
are invariant under the ${\bf Z}_5$, for the simple reason that
by construction the group action acts trivially on the canonical
bundle, and hence $\Lambda^2 {\cal N}_{C/X}$ ({\it i.e.}, the group action
preserves the holomorphic top-form).
Explicitly, we can write a \v{C}ech cocycle on the single overlap patch
(where both $x_1 \neq 0$ and $x_3 \neq 0$) as
\begin{displaymath}
\Omega \: = \: \frac{ x_1 d x_3 \: - \: x_3 d x_1 }{ x_1 x_3 }
\end{displaymath}
which is manifestly ${\bf Z}_5$-invariant. 

Since the gauge bundles are trivial, the spectral sequence
degenerates\footnote{In previous examples, the spectral
sequence degenerated because the cohomology was computed
over a point.  Here, by contrast, the cohomology is not being
computed over a point.  Instead, the gauge bundle is trivial,
so in the language used in the earlier computations,
the evaluation map $\beta$ inside the differential $d_2$
necessarily vanishes, meaning that the boundary conditions
on Neumann $\theta_i$ are not twisted, hence $d_2 = 0$.
More formally, since ${\mathcal E}$ is trivial, its Atiyah class is
automatically $=0$ and so $d_{2} = 0$, yielding the degeneration of
the spectral sequence.}
at $E_{2}$,
and we can now read off the desired Ext groups:
\begin{displaymath}
\mbox{Ext}^p_{ [ X/ {\bf Z}_5 ]} \left( {\cal O}_C, {\cal O}_C \right)
\: = \:
\left\{ \begin{array}{ll}
        {\bf C} & p = 0, \\
        0 & p = 1, 2, \\
        {\bf C} & p=3, \\
        0 & p > 3. 
        \end{array}   \right.
\end{displaymath}
Alternatively, using the results above, it is trivial to compute
the Ext groups on the covering space $X$, following \cite{orig},
and then taking ${\bf Z}_5$-invariants again gives us the result above.

\subsubsection{ ${\bf Z}_5^3$ action on a general quintic threefold}

For another example, consider a more general quintic threefold $X$, defined
by the hypersurface
\begin{displaymath}
x_0^5 \: + \: x_1^5 \: + \: x_2^5 \: + \: x_3^5 \: + \: x_4^5 \: - \:
5 \psi x_0 x_1 x_2 x_3 x_4 \: = \: 0
\end{displaymath}
This has a genus 6 curve $C$ defined by
\begin{displaymath}
x_0^5 + x_1^5 + x_2^5 \: = \: 0, \: x_3 \: = \: x_4 \: = \: 0
\end{displaymath}
whose normal bundle is the restriction of the bundle
${\cal O}(1) \oplus {\cal O}(1)$ to $C$, since $C$ is a complete intersection
of $X$ with the two sections $x_3,\ x_4$  of ${\cal O}(1)$.

This curve is mapped into itself by a ${\bf Z}_5^3$ whose three generators
can be taken to have the following actions on the homogeneous coordinates:
\begin{eqnarray*}
\left( x_0, x_1, x_2, x_3, x_4 \right) & \mapsto &
\left( x_0, \alpha x_1, x_2, x_3, \alpha^4 x_4 \right), \\
\left( x_0, x_1, x_2, x_3, x_4 \right) & \mapsto &
\left( x_0, x_1, \alpha x_2, x_3, \alpha^4 x_4 \right), \\
\left( x_0, x_1, x_2, x_3, x_4 \right) & \mapsto &
\left( x_0, x_1, x_2, \alpha x_3, \alpha^4 x_4 \right).
\end{eqnarray*}
The ${\bf Z}_5$ generated by the third action above leaves the curve
completely invariant, in the sense that it maps every point on the
curve back into itself.  The other two ${\bf Z}_5$'s also map the
curve back into itself, but act nontrivially on the points of the curve.

As before, consider a single D2 brane wrapped on this curve,
with trivial gauge bundle.  In order to define the orbifold
of the theory with the D-brane, we must pick a ${\bf Z}_5^3$-equivariant
structure ({\it i.e.}, a choice of ${\bf Z}_5^3$ action on the Chan-Paton
factors).  For simplicity, we shall take the trivial equivariant structure.

Next, we shall compute $\mbox{Ext}^*_{ [ X/ {\bf Z}_5^3 ] }
\left( {\cal O}_C, {\cal O}_C \right)$.

For this calculation, it suffices to compute the $H^i(C,\Lambda^j
{\cal N}_{C/X})$.  These can only be nonzero if $i=0$ or 1 for
dimension reasons and $j=0,1$, or 2 for rank reasons.  Since
${\cal N}_{C/X}\simeq {\cal O}_C(1) \oplus {\cal O}_C(1)$, we see that
$\Lambda^2{\cal N}_{C/X}\simeq {\cal O}_C(2)$ and so we only need the
cohomology of ${\cal O}_C,\ {\cal O}_C(1)$, and ${\cal O}_C(2)$.
The non-zero cohomology groups are
\[
H^0({\cal O})={\bf C},\ H^1({\cal O})={\bf C}^6,\
H^0({\cal O}_C(1))\simeq H^1({\cal O}_C(1))\simeq
{\bf C}^3,\]
\[
H^0({\cal O}_C(2))\simeq{\bf C}^6,\ H^1({\cal O}_C(2))\simeq{\bf C}.\]
To see the above, note that $H^0({\cal O}_C(1))$ has basis
$x_0,x_1,x_2$; and since $C$ has genus $g=6$ and ${\cal O}_C(1)$ has
degree $d=5$ (the degree of $C$), then $H^1({\cal O}_C(1))\simeq {\bf
C}^3$ follows by Riemann-Roch.  Furthermore,
$H^0({\cal O}_C(2))$ has basis $x_0^2,x_0x_1,x_0x_2,x_1^2,x_1x_2,x_2^2$.
Then $H^1({\cal O}_C(2))\simeq{\bf C}$ follows from Riemann-Roch.  Or note
instead that by adjunction $K_C\simeq{\cal O}_C(2)$ and $H^1(K_C)\simeq
H^{1,1}(C)={\bf C}$.

By the above explicit description, we see that the only ${\bf Z}_5^3$
invariants are $H^0({\cal O})={\bf C}$ and
$H^1(\Lambda^2{\cal N}_{C/X})\simeq H^1({\cal O}_C(2))\simeq{\bf C}
$.  Putting this all together, we get
\begin{displaymath}
\mbox{Ext}^p_{ [ X/ {\bf Z}_5^3 ]} \left( {\cal O}_C, {\cal O}_C \right)
\: = \:
\left\{ \begin{array}{ll}
        {\bf C} & p = 0, 3,\\
        0 & p = 1, 2,\\
        0 & p > 3.
        \end{array}   \right.
\end{displaymath}

\subsubsection{Nontriviality of the spectral sequence}

It is straightforward to combine the techniques used in the last
example with the analysis in \cite{orig} to construct an example in
which the spectral sequence is non-trivial.

We start with $X={\bf P}(1,1,2,2,2)[8]$.  Recall that projection onto the
first two coordinates defines a K3 fibration over ${\bf P}^1$.
We consider the $G={\bf Z}_8$ action on the base of the fibration
\[
g:(x_0,x_1)\mapsto(x_0,\zeta x_1)
\]
where $\zeta = {\rm exp}(2\pi i/8)$.  We assume that this action
preserves the Calabi-Yau.  So for example the equation of the
hypersurface can have the form

\[
x_0^8+x_1^8+f(x_2,x_3,x_4)=0,
\]
where $f$ is a homogeneous polynomial of degree 4 in the variables
$x_2,x_3,x_4$.  It is not difficult to see using standard classical
enumerative algebraic geometry that if $f$ is chosen generically, then
there are finitely many lines contained in the K3 fibers of $X$, each
of which have normal bundle ${\cal O}(-1)\oplus {\cal O}(-1)$.  Pick
one of them and call it $C_0$.  Then $C_i:=g^i\cdot C_0$ is another
such curve.  Since the ${\bf Z}_8$ acts on the base, the $C_i$
are pairwise disjoint.  If we put $C=\cup_iC_i$, then the above computation
of
${\rm Ext}^*_{ [X/{\bf Z}_8]}({\cal O}_C,{\cal O}_C)$
simplifies since there are no intersecting pairs.  Taking the trace as
above, we see that
\[
{\rm Ext}^*_X({\cal O}_C,{\cal O}_C)^G\simeq {\rm Ext}^*_X({\cal O}_{C_0},
{\cal O}_{C_0}).
\]
These latter Ext groups were calculated in \cite{orig}, where it was shown
that the spectral sequence is nontrivial if $C_0$ is chosen generically.
But computing ${\rm Ext}^*({\cal O}_C,{\cal O}_C)$
via the spectral sequence being used in this paper
and then taking invariants reduces to exactly the same computation done
previously in \cite{orig}.  We conclude that the spectral sequence for
this wrapped D-brane on the orbifold of $X$ by $G$ is also non-trivial.

\section{Parallel branes on submanifolds of different dimension}  \label{pardiff}

\subsection{Basic analysis}

The analysis here proceeds much as in the above and in 
\cite{orig}.  Let $S$ be a smooth submanifold as above,
and let $T$ be a submanifold of $S$, also closed under the $G$-action
on $X$.  Let ${\cal E}$, ${\cal F}$ be holomorphic vector bundles
on $S$, $T$, respectively, with $G$-equivariant structures chosen on both.

The massless boundary Ramond sector states can be written in the form
\begin{displaymath}
b^{\alpha \beta j_1 \cdots j_m}_{ 
\overline{\imath}_1 \cdots \overline{\imath}_n}(\phi_0)
\eta^{ \overline{\imath}_1 } \cdots
\eta^{ \overline{\imath}_n }
\theta_{j_1} \cdots \theta_{j_m}
\end{displaymath}
where $\alpha$, $\beta$ are Chan-Paton factors describing
the bundles ${\cal E}|_T$ and ${\cal F}$,
the $\eta$ indices are tangent to $T$,
and (momentarily ignoring the twisting of \cite{abooetal})
the $\theta$ indices are normal to $S \hookrightarrow X$.
As before, to get the orbifold spectrum, one takes the $G$-invariants
of the spectrum on the covering space, so 
naively, the BRST cohomology classes of such states are counted
by the sheaf cohomology groups
\begin{displaymath}
H^n\left(T, {\cal E}^{\vee}|_T \otimes {\cal F} \otimes 
\Lambda^m {\cal N}_{S/X}|_T \right)^G.
\end{displaymath}

As the reader no doubt expects, there is a spectral sequence
\begin{displaymath}
H^n\left(T, \left( {\cal E}|_T \right)^{\vee} \otimes {\cal F}
\otimes \Lambda^m {\cal N}_{S/X}|_T \right)^G \: \Longrightarrow \:
\mbox{Ext}^{n+m}_{ [X/G] }\left( i_* {\cal E}, j_* {\cal F} \right)
\end{displaymath}
(See appendix~\ref{pfs} for a proof.)
In passing, note this spectral sequence can be equivalently written as
\begin{displaymath}
H^n\left( [T/G], \left( {\cal E}|_T \right)^{\vee} \otimes {\cal F}
\otimes \Lambda^m {\cal N}_{S/X}|_T \right) \: \Longrightarrow \:
\mbox{Ext}^{n+m}_{ [X/G] }\left( i_* {\cal E}, j_* {\cal F} \right)
\end{displaymath}
to emphasize the connection with stacks.

In passing, note that we do not need to assume that the
brane configuration is supersymmetric,
because we are counting massless boundary {\it Ramond} sector states.
In non-supersymmetric configurations, that means we are counting
the spectrum of spacetime fermions only.
This same issue, along with analogous issues involving
potential problems with Serre duality, 
appeared in our previous work \cite{orig} and are dealt with in the
same way;
we refer the reader there for more details.

In the simpler cases described in section~\ref{parcoin}, we were able
to show that the spectral sequence appearing there was
realized directly in BRST cohomology, so that the physical
states are counted by Ext groups, and not sheaf cohomology.
We are not able to perform such a direct check in this section;
however,
as in \cite{orig}, we conjecture (based on our experience with
parallel coincident branes) that the spectral sequence of this
section is also realized
physically in BRST cohomology, ultimately because of the Chan-Paton-induced
twisting of the boundary conditions \cite{abooetal}.
Thus, assuming our conjecture is correct, we have again that 
states in the massless boundary Ramond sector spectrum 
are in one-to-one correspondence with elements of the Ext groups
above.

Also note that as before, we have {\it not} assumed that
string orbifolds describe strings propagating on quotient stacks.
Rather, we have merely computed the massless boundary Ramond
spectrum, directly in BCFT, using standard methods,
and noted that, at the end of the day, the spectrum can be
counted by Ext groups on quotient stacks.
We are not assuming any physical relevance of stacks;
rather, we have derived that Ext groups on quotient stacks
count massless boundary Ramond sector states.

Also, just as in \cite{orig}, there is naively a problem
with Serre duality, which can be fixed by assuming the boundary
vacua are sections of certain line bundles.  As the analysis
is completely identical to that in \cite{orig}, we shall
simply refer the reader to \cite{orig} for more information.

\subsection{Example:  ADHM/ALE}

First let us recall the ADHM construction on flat space,
then we shall review the ADHM construction on ALE spaces.
See for example \cite{kronnak,nak} for more detailed
discussion and relevant conventions.

In terms of six-dimensional gauge theory, the moduli space
of $k$ instantons of $U(N)$ on ${\bf R}^4$ is described as the
(classical) Higgs moduli space of the six-dimensional $U(k)$
gauge theory with one adjoint-valued hypermultiplet and
one hypermultiplet valued in the $(k,N)$ of $U(k) \times U(N)$.
More formally, if we let $V = {\bf C}^k$, $W = {\bf C}^N$, 
and $Q = {\bf C}^2$, then the scalars can be described
by the vector space
\begin{displaymath}
\mbox{Hom}\left(V, Q \otimes V\right) \oplus
\mbox{Hom}\left(W,V\right) \oplus
\mbox{Hom}\left(V,W\right).
\end{displaymath}

The ADHM construction on an ALE space is closely related.
In addition to the orbifold group action on the base space ${\bf R}^4$,
we must also define orbifold group actions on the gauge bundle
and Higgs fields.  In the more formal language above, that
means that in addition to the orbifold group action on $Q$
($Q$ is identified with ${\bf C}^2 = {\bf R}^4$, and hence
automatically has an orbifold group action defined), we must
also pick orbifold group actions on $V$ and $W$.  
The matter fields in the six-dimensional gauge theory are those
of the original ADHM construction that are left invariant
under the $G$ action.  In other words, in the language above,
the remaining matter fields are identified with the $G$-invariant
vector space homomorphisms
\begin{displaymath}
\mbox{Hom}_G\left(V, Q \otimes V\right) \oplus
\mbox{Hom}_G\left(W,V\right) \oplus
\mbox{Hom}_G\left(V,W\right).
\end{displaymath}

To compare to our description of massless modes,
consider $N$ D4-branes (with trivial gauge bundle)
on ${\bf R}^4$, with an orbifold
group action $G: {\bf R}^4 \mapsto {\bf R}^4$ defining the
ALE quotient.  Consider $k$ D0-branes (with trivial gauge bundle)
located at the origin
of ${\bf R}^4$, {\it i.e.}, at the fixed point of $G$.
The worldvolumes of the branes are invariant under the group action,
as needed.  As described earlier, whenever one has a gauge field,
one must specify an orbifold group action on the gauge field -- 
specifying an orbifold group action on the base space does not
uniquely define an orbifold group action on the gauge field,
as one could always combine the orbifold group action with a 
gauge transformation.  In the present case, for trivial bundles,
specifying a $G$-action on a trivial rank $N$ bundle is equivalent
to specifying an $N$-dimensional representation of $G$.

We described the massless Ramond sector states for
$k$ D0-branes on $N$ D4-branes in this context in \cite{orig}.
There, we worked through how our methods recover the standard
ADHM construction.  In the present case, specifying $G$-equivariant
structures on the bundles is equivalent to specifying
$k$-dimensional and $N$-dimensional representations of $G$,
and then the massless spectrum is the part of the original ADHM
spectrum that is invariant under the action of $G$.

More technically, in the unorbifolded theory, from open strings
connecting the D0 and D4 branes, we recovered \cite{orig}
\begin{eqnarray*}
\mbox{Ext}^n_{ {\bf C}^2 }\left(i_* {\cal E}, j_* {\cal F} \right) & = & 
\left\{ \begin{array}{ll}
        {\bf C}^{kN} = \mbox{Hom}(W,V) & n = 0, \\
        0 & n \neq 0, 
        \end{array} \right.  \\
\mbox{Ext}^n_{ {\bf C}^2 }\left(j_* {\cal F}, i_* {\cal E} \right) & = & 
\left\{ \begin{array}{ll}
        {\bf C}^{kN} = \mbox{Hom}(V,W) & n = 2, \\
        0 & n \neq 2. 
        \end{array} \right.  
\end{eqnarray*}
In the present orbifolded theory, from open strings connecting
the D0 and D4 branes, we recover
\begin{eqnarray*}
\mbox{Ext}^n_{ [{\bf C}^2/G] }\left( i_* {\cal E}, j_* {\cal F} \right)
& = &
\left\{ \begin{array}{ll}
        ({\bf C}^{kN})^G = \mbox{Hom}_G(W,V) & n=0, \\
        0 & n \neq 0,
        \end{array} \right. \\
\mbox{Ext}^n_{ [{\bf C}^2/G] }\left( j_* {\cal F}, i_* {\cal E} \right)
& = &
\left\{ \begin{array}{ll}
        ({\bf C}^{kN})^G = \mbox{Hom}_G(V,W) & n=2, \\
        0 & n \neq 2. 
        \end{array} \right. 
\end{eqnarray*}
Also as in \cite{orig}, we see that the matter fields arise
from Ext groups of degree zero.\footnote{Note that we have extended
supersymmetry in this example.}

In short, we immediately recover the ADHM construction on ALE spaces.

\section{General intersections}   \label{gencase}

The analysis for general intersections
proceeds much as in the above and in \cite{orig}.
Let $S$, $T$ be smooth submanifolds, both closed under the $G$ action
on $X$, and let ${\cal E}$, ${\cal F}$ be holomorphic vector 
bundles on $S$, $T$, respectively, both with $G$-equivariant
structures chosen.  In order to write down the massless
boundary Ramond sector states for the general case, we need
to take into account facts about branes at angles \cite{bdl}.
We refer the reader to \cite{orig} for details of the analysis
on the covering space.
Suffice it to say that the massless boundary Ramond sector spectrum,
ignoring subtleties in boundary conditions, appears to be counted by the sheaf
cohomology groups
\begin{displaymath}
\begin{array}{c}
H^p\left( S \cap T, {\cal E}^{\vee}|_{S \cap T} \otimes
{\cal F}|_{S \cap T} \otimes \Lambda^{q-m} \tilde{N} \otimes
\Lambda^{top} {\cal N}_{S \cap T/T} \right)^G, \\
H^p\left( S \cap T, {\cal E}|_{S \cap T} \otimes {\cal F}^{\vee}|_{
S \cap T} \otimes \Lambda^{q-n} \tilde{N} \otimes \Lambda^{top} {\cal N}_{S \cap T/S}
\right)^G. 
\end{array}
\end{displaymath}
Just as in \cite{orig},
the line bundles $\Lambda^{top} {\cal N}_{S \cap T/T}$
and $\Lambda^{top} {\cal N}_{S \cap T/S}$ are a reflection of the
Freed-Witten anomaly \cite{fw}, which plays an essential role.
As discussed in \cite{orig,fw,lec}, in order to make the Chan-Paton factors
well-defined if the normal bundle does not admit a Spin structure,
the Chan-Paton factors must be twisted, and so the sheaf $i_* {\cal E}$
actually corresponds to the ``bundle'' ${\cal E} \otimes \sqrt{ K_S^{\vee}}$
on the D-brane worldvolume.
Our calculations so far are unaffected, but general intersections pick
up a factor of $\sqrt{ K_S / K_T }$.
Using the fact that a separate anomaly implies that the open string
B model is only well-defined when the product
\begin{displaymath}
\Lambda^{top} {\cal N}_{S \cap T / S} \otimes \Lambda^{top}
{\cal N}_{S \cap T / T}
\end{displaymath}
is trivializable,
it is easy to compute that when the open string B model is well-defined,
\begin{eqnarray*}
\sqrt{ \frac{ K_S|_{S \cap T} }{ K_T|_{S \cap T} } } & = &
\Lambda^{top} {\cal N}_{S \cap T / T } \\
\sqrt{ \frac{ K_T|_{S \cap T} }{ K_S|_{S \cap T} } } & = &
\Lambda^{top} {\cal N}_{S \cap T / S }
\end{eqnarray*}
giving the line bundles above. 
These subtleties are discussed in great detail in
\cite{orig}; we refer the reader there for more information.

As the reader should expect from \cite{orig}, there are
two spectral sequences
\begin{eqnarray*}
H^p\left( S \cap T, {\cal E}^{\vee}|_{S \cap T} \otimes
{\cal F}|_{S \cap T} \otimes \Lambda^{q-m} \tilde{N} \otimes
\Lambda^{top} {\cal N}_{S \cap T/T} \right)^G
& \Longrightarrow & 
\mbox{Ext}^{p+q}_{ [X/G] }\left( i_* {\cal E}, j_* {\cal F} \right), \\
H^p\left( S \cap T, {\cal E}|_{S \cap T} \otimes {\cal F}^{\vee}|_{
S \cap T} \otimes \Lambda^{q-n} \tilde{N} \otimes \Lambda^{top} {\cal N}_{S \cap T/S}
\right)^G 
& \Longrightarrow & \mbox{Ext}^{p+q}_{ [X/G] } \left( j_* {\cal F},
i_* {\cal E} \right),
\end{eqnarray*}
where 
\begin{displaymath}
\tilde{N} \: = \: TX|_{S \cap T} / \left( TS|_{S \cap T} + TT|_{S \cap T} \right)
\end{displaymath}
and $m = \mbox{rk } {\cal N}_{S \cap T/T}$, $n = \mbox{rk } {\cal N}_{S \cap T/S}$.
(See appendix~\ref{pfs} for a proof.)
In passing, note these spectral sequences can be equivalently written as
\begin{eqnarray*}
H^p\left( [ S \cap T / G],  {\cal E}^{\vee}|_{S \cap T} \otimes
{\cal F}|_{S \cap T} \otimes \Lambda^{q-m} \tilde{N} \otimes
\Lambda^{top} {\cal N}_{S \cap T/T} \right) & \Longrightarrow &
\mbox{Ext}^{p+q}_{ [X/G] }\left( i_* {\cal E}, j_* {\cal F} \right), \\
H^p\left( [S \cap T/G],  {\cal E}|_{S \cap T} \otimes {\cal F}^{\vee}|_{
S \cap T} \otimes \Lambda^{q-n} \tilde{N} \otimes \Lambda^{top} {\cal N}_{S \cap T/S}
\right) & \Longrightarrow &
 \mbox{Ext}^{p+q}_{ [X/G] } \left( j_* {\cal F},
i_* {\cal E} \right),
\end{eqnarray*}
to emphasize the connection with stacks.

All the subtleties that arose in \cite{orig} have immediate analogues here.
For purposes of brevity, we merely refer the reader to \cite{orig} for
more details.  Just as in \cite{orig}, we conjecture that both
of the spectral sequences above are realized physically in BRST
cohomology, due ultimately to Chan-Paton-induced boundary condition
twistings of the form described in \cite{abooetal},
so that the massless boundary Ramond spectrum is calculated by
Ext groups.

It is also worth repeating that, as before, we are {\it not}
assuming any physical relevance of quotient stacks.
Rather, we have done a first-principles calculation of 
massless boundary Ramond sector states, directly in BCFT,
and found that, at the end of the day, those states are counted
by Ext groups on quotient stacks.

\section{Relating orbifolds to large radius, or, the McKay correspondence} \label{mckay}

For the purposes of understanding monodromy in K\"ahler moduli space,
it is useful to have a correspondence between sheaves on orbifolds
and sheaves on large radius resolutions.
Such a correspondence exists, and is known as the McKay correspondence.
In the form most relevant to this paper, the McKay correspondence
says that \cite{mimpm}
\begin{displaymath}
D_G\left(X\right) \: \cong \: D\left( \widetilde{X/G} \right)
\end{displaymath}
{\it i.e.}, the bounded derived category of $G$-equivariant sheaves
on $X$ is isomorphic to the bounded derived category of
sheaves on a resolution $\widetilde{X/G}$.

We can also translate this statement into quotient stacks very quickly.
The bounded derived category of $G$-equivariant sheaves on $X$ is the
same thing as the bounded derived category of sheaves on
the quotient stack $[X/G]$, {\it i.e.},
\begin{displaymath}
D\left( [X/G] \right) \: = \:
D_G\left( X \right),
\end{displaymath}
so we can rewrite the McKay correspondence as the statement that
\begin{displaymath}
D\left( [X/G] \right) \: \cong \: D\left( \widetilde{ X/G } \right).
\end{displaymath}

Written in the form above, the McKay correspondence can immediately
be used to relate open string spectrum computations at large radius
to the open string spectrum computations in orbifolds that we have
performed in this paper.  Specifically,  
let $DF: D(
\widetilde{X/G}) \rightarrow D([X/G])$ 
denote the (derived) functor defining the equivalence
of (derived) categories.  Let ${\cal E}$, ${\cal F}$ be two
sheaves on $\widetilde{X/G}$.
Then
\begin{displaymath}
\mbox{Ext}^n_{ \widetilde{X/G} }\left( {\cal E}, {\cal F} \right)
\: = \:
\mbox{Ext}^n_{ [X/G] }\left( DF( {\cal E}),
DF( {\cal F} ) \right) 
\end{displaymath}
In other words, the McKay correspondence preserves Ext groups.

Note that we are simply identifying a set of branes at large radius
with a set of branes at the orbifold point, using the McKay
correspondence, and are making no claims about
any physical flow of branes with K\"ahler moduli.

\subsection{Images of fractional D0 branes} \label{ss-fracD0}

As an example of the McKay correspondence formalism we compute the
large radius limit of an irreducible fractional D0 brane in $[ {\bf
C}^3 / {\bf Z}_3 ]$. To make the notation less cumbersome we write $X
= {\bf C}^{3}$,  $G = {\bf Z}_{3}$. Let $M = X/G$ denote the quotient
as a singular variety and let $Y = \widetilde{M} = \widetilde{X/G}$ be
the canoincal crepant resolution of $M$. Explicitly we choose
coordinates $x, y, z$ on $X = {\bf C}^{3}$ in terms of which the
action of the generator in   $G =
{\bf Z}_{3}$ is given by $(x,y,z) \mapsto (\mu x,\mu y, \mu z)$, where
$\mu = \exp(2\pi i/3)$ is a fixed third root of unity. The space $Y$ is
the blow up of the unique singular point of $M$. It can be described
explicitly as follows. Write $P = {\bf P}({\mathbb C}^{3})$ for the
projective plane with homogeneous coordinates $(x,y,z)$. Then $Y =
{\rm tot}({\mathcal O}_{P}(-3))$ is the total space of the line bundle
${\mathcal O}_{P}(-3)$ and the natural map $\tau : Y \to M$
is simply contracting the zero section. Let $Z$ denote the fiber
product of $Y$ and $X$ over $M$. Then $Z$ can be identified with the
total space $Z = {\rm tot}({\mathcal O}_{P}(-1))$ and the map $q : Z
\to X$ is again the contraction of the zero section, this time to a
smooth point - the origin $o = (0,0,0) \in X = {\bf C}^{3}$. All this
data can be conveniently organized in the commutative diagram
\[
\xymatrix{
& Z \ar[rr]^-{q} \ar[dd]_-{p} \ar[dl]_-{\zeta} & & X \ar[dd]^-{\pi} \\
P & & & \\
& Y \ar[rr]_-{\tau} \ar[ul]^-{\eta} & & M 
}
\]
where $\eta : Y \to P$ and $\zeta : Z \to P$ denote the natural
projections and $p : Z \to Y$ and $\pi : X \to M$ are the maps of
taking a quotient by $G$. Note that the group $G$ acts on $Z$ by
simply multiplying by $\mu$ along the fibers of ${\mathcal O}_{P}(-1)
\to P$ and so the map $p : Z \to Y$ can also be viewed as the map of
raising into third power along the fibers of the line bundle
${\mathcal O}_{P}(-1)$. Conversely, we can view $Z$ as the canonical
cubic root cover of $Y$ which is branched along the zero section $B
\subset Y$ of $\eta$. For future reference we denote  the zero section
of $\zeta$ by $R \subset Z$. With this notation we are ready to write the
explicit formula defining the McKay correspondence $DF$. Indeed,
following \cite{mimpm} we have that $DF$ is given by:
\[
\xymatrix@R-9pt{
DF : & D^{b}(Y) \ar[r] & D^{b}_{G}(X) \\
& {\mathcal F} \ar@{{|}->}[r] & Rq_{*}p^{*}{\mathcal F},
}
\]
where the object\footnote{In principle one should take here the
derived pullback $Lp^{*}{\mathcal F}$. However the map $p : Z \to Y$
is finite and flat and so $Lp^{*} = p^{*}$.} $p^{*}{\mathcal F}$ on
$Z$ is equipped with the trivial $G$-equivariant structure.

Let us examine now what becomes of the fractional D0 branes on the
stack quotient $[X/G]$ when we transform them back to the resolution
$Y$ via the inverse McKay correspondence 
\[
DF^{-1} : D^{b}([X/G]) =
D^{b}_{G}(X) \to D^{b}(Y).
\] 
As we already explained, a sheaf on $[X/G]$ will
represent a fractional D0 brane if it is
supported at the gerby point $[o/G] = BG \hookrightarrow [X/G]$.
Equivalently, a set of fractional D0 brane on $[X/G]$ can be thought of as an
object in $D^{b}_{G}(X)$ whose cohomology sheaves are $G$-equivariant
sheaves on $X$ whose reduced support is the point $o \in X$. For
concreteness we will discuss only pure fractional D0 branes. For these
the underlying object in $D^{b}(X)$ is just the structure sheaf
${\mathcal O}_{o}$ of the reduced point $o \in X$ (extended by zero to
a sheaf on $X$). Note that the ideal sheaf $I_{o} \subset {\mathcal
O}_{X}$ of the point $o$ has a natural equivariant structure and so
${\mathcal O}_{o}$ inherits a natural $G$-equivariant
structure. Specifying another equivariant structure on ${\mathcal
O}_{o}$ amounts to simply multiplying the natural one by a a character
$\alpha: G \to {\bf C}^{*}$. We will denote the corresponding
$G$-equivariant sheaf by $({\mathcal O}_{o},\alpha)$. Let $\chi : G
\to {\bf C}^{*}$ be the character sending $1 \in {\mathbb Z}_{3}$ to
$\mu \in {\bf C}^{3}$. Then the group of characters of $G$ consists of
the trivial character $\boldsymbol{1}$ and the two non-trivial
characters $\chi$ and $\chi^{2}$ and so we need to compute the objects 
$DF^{-1}(({\mathcal O}_{o},\boldsymbol{1}))$, $DF^{-1}(({\mathcal
O}_{o},\chi))$ and $DF^{-1}(({\mathcal O}_{o},\chi^{2}))$ in
$D^{b}(Y)$. 

Taking into account that the varieties $X$, $Y$ and $Z$ are all smooth
and using the relative duality formula we see that $DF^{-1}$ is given
as ${\mathcal G} \to (Rp_{*}q^{!}{\mathcal G})^{G} =
(Rp_{*}(Lq^{*}{\mathcal G}\otimes^{L} \omega_{Z/X}))^{G}$, where $\omega_{Z/X}$
denotes the relative dualizing sheaf of the map $q : Z \to X$. Next
observe, that the map $p : Z \to Y$ is affine and so $Rp_{*} =
p_{*}$. Thus computing $DF^{-1}{\mathcal G}$ of an equivariant complex
${\mathcal G}$ on $X$, reduces essentially to computing the derived
pullback $Lq^{*}{\mathcal G}$ with the induced $G$-equivariant
structure. For this, we only need to note that the the sheaf
${\mathcal O}_{o}$ on $X$ admits the standard Koszul resolution:
\[
\xymatrix@1@C+15pt{
0 \ar[r] & {\mathcal O}_{X} \ar[rr]^-{\begin{pmatrix} x & y & z
\end{pmatrix}} & & {\mathcal O}_{X}^{\oplus 3} 
\ar[rr]^*+[u]{\txt{{\small $\left(\begin{array}{ccc} 0 &-z & y \\ z & 0 &
-x \\ -y & x & 0 
\end{array}\right)$}}}  
 & & {\mathcal O}_{X}^{\oplus 3} \ar[rr]^-{\begin{pmatrix} x \\ y \\ z
\end{pmatrix}} & & {\mathcal O}_{X} \ar[r] & {\mathcal O}_{o},
}
\]
and so $Lq^{*}{\mathcal O}_{o}$ is quasi-isomorphic to the complex 
\begin{equation} \label{eq-Lq}
\xymatrix@1@C+15pt{
{\mathcal O}_{Z} \ar[rr]^-{\begin{pmatrix} x & y & z\end{pmatrix}} & &
{\mathcal O}_{Z}^{\oplus 3} 
\ar[rr]^*+[u]{\txt{{\small $\left(\begin{array}{ccc} 0 & -z & y \\ z & 0 &
-x \\ -y & x & 0 
\end{array}\right)$}}}
 & & {\mathcal O}_{Z}^{\oplus 3} \ar[rr]^-{\begin{pmatrix} x \\ y \\ z
\end{pmatrix}} & & {\mathcal O}_{Z},
}
\end{equation}
where the last ${\mathcal O}_{Z}$ on the right is placed in degree $0$
and we have used the same notation $x$, $y$ and $z$ for the pullbacks
of the coordinate functions on $X = {\bf C}^{3}$ to $Z$. Now it is
clear that the cohomology sheaves of the complex \eqref{eq-Lq} are all
supported on $R$, and by inspecting the maps in \eqref{eq-Lq} we see
that the cohomology sheaves of $Lq^{*}{\mathcal O}_{o}$ are
\[
L^{0}q^{*}{\mathcal O}_{o} = {\mathcal O}_{R}, \qquad 
L^{-1}q^{*}{\mathcal O}_{o} = \Omega^{1}_{R}(1), \qquad
L^{-2}q^{*}{\mathcal O}_{o} = \Omega^{2}_{R}(2).
\]
Furthermore, taking into account that $q : Z \to X$ is a blow-up of
the point $o$ with exceptional divisor $R \subset Z$ we see that
$\omega_{Z/X} = {\mathcal O}_{Z}(2R) = \zeta^{*}{\mathcal
O}_{P}(-2)$. 
In particular, if we  trace out the equivariant structures we see that
\[
\begin{split}
L^{0}q^{*}({\mathcal O}_{o},\boldsymbol{1})\otimes \omega_{Z/X} & = 
({\mathcal O}_{R}(-2),\chi^{2}),  \\
L^{-1}q^{*}({\mathcal O}_{o},\boldsymbol{1})\otimes \omega_{Z/X} &  =
(\Omega^{1}_{R}(-1),\chi), \\
L^{-2}q^{*}({\mathcal O}_{o},\boldsymbol{1})\otimes \omega_{Z/X} & =
(\Omega^{2}_{R},\boldsymbol{1}),
\end{split}
\]
as $G$-equivariant sheaves on $Z$. Therefore, if we pushforward by $p$
and take $G$-invariants we get the following images of the fractional
D0 branes:
\[
\begin{split}
DF^{-1}({\mathcal O}_{o},\boldsymbol{1}) & = \Omega_{B}^{2}[2] \\
DF^{-1}({\mathcal O}_{o},\chi) & = {\mathcal O}_{B}(-2) \\
DF^{-1}({\mathcal O}_{o},\chi^{2}) & = \Omega_{B}^{1}(-1)[1].
\end{split}
\]
\
These results were also obtained in \cite{ci}.

\subsection{Images of D0 branes on resolutions}

It is also instructive to examine the McKay image of a skyscraper 
sheaf ${\mathcal O}_{b}$ on $Y$ supported at a reduced point $b \in B 
\subset Y$. Consider the image $\eta(b) \in P$ of $b$ under the 
projection map $\eta : Y \to P$ and let $(u,v)$ be linear affine 
coordinates on $P$ around the points $\eta(b)$. Let $t$ and $s$ be 
linear coordinates along the fibers of $\eta$ and $\zeta$ 
respectively. Then locally near $b$ the variety $Y$ has a chart with 
coordinates $(u,v,t)$, locally near the point $p^{-1}(b) \in R \subset 
Z$, the variety $Z$ has coordinates $(u,v,s)$ and the points $b$ and 
$p^{-1}(b)$ are given by $u=v=t=0$ and $u=v=s=0$ 
respectively. Furthermore, in these terms the map $p : Z \to Y$ is 
given by $(u,v,s) \mapsto (u,v,s^{3}) = (u,v,t)$ and so 
$p^{*}{\mathcal O}_{b}$ is the non-reduced scheme on $Z$, which is 
supported at the point $p^{-1}(b)$ and is cut out by the ideal 
$\langle u, v, s^{3} \rangle$. Pushing forward by $q$ results in a 
$G$-equivariant sheaf $DF({\mathcal O}_{b})$ which is the structure 
sheaf of the unique triple structure on the point $o \in X$ which is a 
closed subscheme of the line $L_{b} \subset X = {\bf 
C}^{3}$ corresponding to the point $b \in P = {\bf P}({\mathbb 
C}^{3})$. Note that the line $L_{b}$ is naturally stable under the 
$G$-action on $X$ and so the torsion sheaf $DF({\mathcal O}_{b}) = 
{\mathcal O}_{L_{b}}/{\mathcal O}_{L_{b}}(-3o)$ has a canonical 
structure of a $G$-equivariant sheaf. 

Physically, such non-reduced schemes are suspected to be
equivalent to D-branes with nonzero nilpotent Higgs fields
\cite{tomasme}.  Such a relationship is implicit in some of
the older work on D-brane moduli spaces, where authors used to
sometimes claim that D-branes appeared to resolve singularities
(see for example \cite{dgm}).
The Higgs vacua seen were typically nilpotent Higgs vevs assigned
to D-branes on the orbifold, hence described mathematically
by non-reduced schemes, and by virtue of the McKay correspondence,
we see here explicitly they are related to large-radius resolutions.
From this perspective, the resolutions described in \cite{dgm} are
really non-reduced schemes on the quotient stack; the apparent
relation to resolutions is a consequence of the McKay correspondence.

\section{Conclusions}  \label{conclusions}

In this paper we have performed a first-principles calculation,
directly in BCFT,
of the massless boundary Ramond sector spectrum in open strings
in orbifolds, and discovered that that spectrum is counted by
Ext groups on quotient stacks, just as the same spectrum
for open strings on smooth Calabi-Yau's was counted by Ext groups
on the Calabi-Yau's \cite{orig}.

Although the technical statement above sounds complicated,
it really comes from a simple idea, first stated in
\cite{dougmoore}, namely that one can compute open string
spectra in orbifolds by first computing spectra on the covering
space, and then taking group invariants.
Here, the relevant spectrum on the covering space is counted
by elements of Ext groups on the covering space,
and the group invariants of the Ext groups on the cover
are the Ext groups on the quotient stack.
Though the detailed verification that this is consistent
with the manner in which Ext groups are realized physically takes work,
the intuition is just this simple.

At no point in this analysis did we assume that string orbifolds
are the same as strings compactified on quotient stacks.
However, our results are certainly evidence for such a proposal.
Mathematically, quotient stacks are
a tool for manipulating orbifolds, but they have all of the underpinnings
necessary to be able to propagate strings on them.
Not only can one define sheaves on stacks, as we have discussed
in this paper, but one can also do differential geometry on stacks,
and in fact most topological notions that physicists are familiar
with carry over to stacks -- stacks really are the next best
thing to spaces.  So, can strings propagate on stacks,
and is a string orbifold the same thing as a string propagating
on a quotient stack?  If so, not only would we gain new physical insight
into string orbifolds, but we would also gain a new class of string
compactifications -- on stacks.

In previous work \cite{meqs}, a proposal
was made for a classical action for sigma models on stacks
that generalizes sigma models on ordinary spaces,
another prerequisite for making sense of the notion of
strings propagating on stacks.  For example,
twisted sectors emerge very naturally when thinking about
sigma models on quotient stacks.  In the same work,
physical characteristics of string orbifold CFT's were also
related to geometric features of stacks.
In other work \cite{agv}, it has been shown that certain
correlation functions in string orbifolds can be rewritten,
using stacks, in a form that is closely analogous to 
correlation function computations on ordinary spaces.
In this paper, by showing that open string spectra in orbifolds are counted
by Ext groups on stacks, we are not only contributing
to attempts to understand the role derived categories play
in physics, but we are also uncovering evidence for the proposal
that string orbifolds coincide with strings propagating on quotient
stacks.

One of the fundamental issues that needs to be addressed
in this area is mathematical, not physical.
Specifically, we need a better understanding of the
deformation theory of stacks.  Deformations of the conformal
field theory underlying string orbifolds have been studied
in the physics literature, and the places occupied by
string orbifolds in CFT moduli spaces are well-understood.
Are quotient stacks consistent with that knowledge of CFT?

\section{Acknowledgements}

We would like to thank D.-E.~Diaconescu  
for useful conversations, A.~Craw and A.~Ishii for pointing out an error
in an earlier version of this paper, and A.~Greenspoon for numerous comments
on a rough draft.  S.K. was partially supported by NSF grant DMS 02-96154 and 
NSA grant MDA904-02-1-0024, and T.P. was partially supported by
NSF grant DMS 0099715, NSF FRG grant DMS 0139799 and an A.P.Sloan 
research fellowship.

\appendix

\section{Notes on stacks}    \label{stxap}

In this appendix we shall give a very slightly more detailed
picture of stacks in general, and quotient stacks in particular.
We shall outline some definitions, and briefly describe how one defines
functions, differential forms, and bundles on stacks,
to give the reader a taste of how this technology works.
For more information, see for example 
\cite{meqs,vistoli,gomez,cm,fantechi,lmb}.
Although several of those references describe only the
algebraic geometry of stacks, one can also do differential
geometry on stacks, and in this appendix, we shall outline
topology and differential geometry, not algebraic geometry, on stacks.

Stacks are defined by the category of incoming continuous
maps, a definition with obvious applications to sigma models.  
In principle, we can define topological spaces
in the same way -- if we know all the continuous maps into a space
$Y$ from all other spaces $X$, we can reconstruct $Y$.
For example, the points of $Y$ are the same as the maps from
a single point into $Y$.
Although we shall not do so here, it is not difficult to check
that the category of continuous maps into a topological space
determines the space.  In particular, since one can define
a topological space in terms of a category of incoming continuous
maps, stacks generalize spaces.
Also note this is analogous to noncommutative spaces,
where one first defines spaces in terms of algebras of functions on them,
and them constructs noncommutative spaces via considering
more general algebras.  Here, instead of working with functions on
the space, we are working with the continuous maps into the space.

As mentioned above, to define a stack, we specify a category that 
defines the incoming 
continuous maps.
For example, the quotient stack
$[X/G]$ is defined by the category whose
\begin{enumerate}
\item objects are pairs $( E \longrightarrow Y, E \stackrel{f}{\longrightarrow}
X)$, where $Y$ is any topological space, $E \rightarrow Y$ is a principal
$G$-bundle, and $f: E \rightarrow X$ is a continuous $G$-equivariant map.
\item morphisms
\begin{displaymath}
\left( \, E_1 \longrightarrow Y_1, \: E_1 \stackrel{f_1}{\longrightarrow} X
\, \right) \: \longrightarrow \:
\left( \, E_2 \longrightarrow Y_2, \: E_2 \stackrel{f_2}{\longrightarrow} X
\, \right)
\end{displaymath}
are pairs $(\rho, \lambda)$, where $\rho: Y_1 \rightarrow Y_2$ is a continuous
map and $\lambda: E_1 \rightarrow E_2$ is a map of principal $G$-bundles,
making certain diagrams commute.
\end{enumerate}

There is a natural map $X \rightarrow [X/G]$,
defined as above by the trivial principal $G$-bundle $X \times G$ on $X$
and the ($G$-equivariant) evaluation map $X \times G \rightarrow X$.
There is also a natural projection map $[X/G] \rightarrow X/G$.
In the special case that $G$ acts freely, $[X/G] \cong X/G$.

Many standard topological notions, such as open maps, closed maps,
local homeomorphisms, and so forth, have analogues for stacks.
A stack is smooth (and one can define differential forms, {\it etc})
if it admits a smooth atlas.  A quotient stack $[X/G]$ is smooth
if and only if $X$ is smooth and $G$ acts by diffeomorphisms.
(Note that $[X/G]$ can be smooth even if the quotient space $X/G$
is singular.)
We do not have the space here to work through such notions in detail;
see instead \cite{meqs,vistoli,gomez} for readable introductions.

To give a little more insight, let us take a moment to define
functions on stacks.
The general idea is that functions, metrics, differential forms, sheaves,
{\it etc} are defined in terms of their pullbacks to other spaces.
If one knows all the possible pullbacks, under all possible maps,
then one knows the original function, metric, differential form, sheaf,
{\it etc}.

To define a real-valued function on a stack ${\cal F}$, 
we associate a function $f_Y: Y \rightarrow {\bf R}$
to each map $Y \rightarrow {\cal F}$ (more precisely, to each
object in the category defining ${\cal F}$),
such that for any $g: Y_1 \rightarrow Y_2$ commuting with the maps
into ${\cal F}$ ({\it i.e.} for each morphism in the category
defining ${\cal F}$), 
\begin{displaymath}
\xymatrix{
Y_1 \ar[rr]^{g} \ar[dr] & & Y_2 \ar[dl] \\
& {\cal F} &
}
\end{displaymath}
one has $f_2 \circ g = f_1$.

In the case that the stack ${\cal F}$ is actually an ordinary space,
this definition uniquely specifies a function on the space
 -- just take the real-valued function associated to the identity
map $X \stackrel{=}{\longrightarrow} X$.  In fact, we could define
functions on ordinary spaces in the same way, in terms of their pullbacks
under all incoming continuous maps.

In the case that ${\cal F} = [X/G]$, a function on ${\cal F}$ is the
same as a $G$-invariant function on $X$.
First, note that since a function on the stack associates a real-valued
function to each space mapping into the stack,
we immediately have a function on the space $X$, since
there is a map $X \rightarrow [X/G]$.
To show that that map is $G$-invariant, we use the fact that
for any $g \in G$,
the map $g: X \rightarrow X$ commutes with the map
$X \rightarrow [X/G]$:
\begin{displaymath}
\xymatrix{
X \ar[rr]^{g} \ar[dr] & & X \ar[dl] \\
& [X/G] & 
}
\end{displaymath}
According to the constraint above,
that means that $f \circ g = f$ for all $g \in G$, {\it i.e.},
$f$ is $G$-invariant.
Conversely, given a $G$-invariant function on $X$,
one can also construct real-valued functions on all other
spaces with maps into $[X/G]$, though we shall not work through
those details here.

We can define other objects on stacks in exactly the same
manner, in terms of their pullbacks to other spaces.
For example, a bundle on an ordinary space $X$
can be defined\footnote{This is not the usual definition
that physicists typically run into, but it is equivalent
to the usual definition.} by the following data:
\begin{enumerate}
\item For all spaces $Y$ and continuous maps $f: Y \rightarrow X$,
a bundle ${\cal S}_f$ on $Y$.
\item For each commuting diagram
\begin{displaymath}
\xymatrix{
Y_1 \ar[rr]^{\rho} \ar[dr]_{f_1} & & Y_2 \ar[dl]^{f_2} \\
& X &
}
\end{displaymath}
an isomorphism $\varphi_{\rho}: {\cal S}_{f_1} \stackrel{\sim}{\longrightarrow}
\rho^* {\cal S}_{f_2}$
which obeys the consistency condition that for all commuting triples
\begin{displaymath}
\xymatrix{
Y_1 \ar[r]^{\rho_1} \ar[dr]_{f_1} &
Y_2 \ar[r]^{\rho_2} \ar[d]^{f_2} &
Y_3 \ar[dl]^{f_3} \\
& X &
}
\end{displaymath}
the isomorphisms obey
\begin{displaymath}
\varphi_{\rho_2 \circ \rho_1} \: = \:
\rho_1^* \varphi_{\rho_2} \circ \varphi_{\rho_1}: \:
{\cal S}_{f_1} \: \stackrel{\sim}{\longrightarrow} \:
(\rho_2 \circ \rho_1)^* {\cal S}_{f_3}
\end{displaymath}
\end{enumerate}

Now, as the reader may have guessed, one defines a bundle on
a stack ${\cal F}$ in close analogy, using the following
data:
\begin{enumerate}
\item For all spaces $Y$ and maps $f: Y \rightarrow {\cal F}$, 
a bundle ${\cal S}_f$ on $Y$.
\item For each commuting diagram
\begin{displaymath}
\xymatrix{
Y_1 \ar[rr]^{\rho} \ar[dr]_{f_1} & & Y_2 \ar[dl]^{f_2} \\
& {\cal F} &
}
\end{displaymath}
an isomorphism $\varphi_{\rho}: {\cal S}_{f_1} \stackrel{\sim}{
\longrightarrow} \rho^* S_{f_2}$, 
which obeys the consistency condition that for all commuting triples
\begin{displaymath}
\xymatrix{
Y_1 \ar[r]^{\rho_1} \ar[dr]_{f_1} &
Y_2 \ar[r]^{\rho_2} \ar[d]^{f_2} &
Y_3 \ar[dl]^{f_3} \\
& X &
}
\end{displaymath}
the isomorphisms obey
\begin{displaymath}
\varphi_{\rho_2 \circ \rho_1} \: = \:
\rho_1^* \varphi_{\rho_2} \circ \varphi_{\rho_1}: \:
{\cal S}_{f_1} \: \stackrel{\sim}{\longrightarrow} \:
(\rho_2 \circ \rho_1)^* {\cal S}_{f_3}
\end{displaymath}
(the same consistency condition that appeared for spaces).
\end{enumerate}
The definition above of a bundle on a stack
clearly specializes to ordinary bundles in the case that the stack
is a space.
 
In the special case that the stack ${\cal F}$ is a quotient
stack $[X/G]$, the data above is equivalent to a $G$-equivariant
bundle on $X$. 
First, since there is a map $X \rightarrow [X/G]$,
a bundle on $[X/G]$ defines a bundle on $X$.
The $G$-equivariant structure on that bundle on $X$ is
a consequence of the isomorphisms induced by the commuting diagrams
\begin{displaymath}
\xymatrix{
X \ar[rr]^{g} \ar[dr] & & X \ar[dl] \\
& [X/G] &
}
\end{displaymath}
that exist for all $g \in G$.
Conversely, given a $G$-equivariant bundle on $X$,
one can define a sheaf on all spaces with maps into $[X/G]$,
satisfying the consistency conditions above.
(Showing this in detail requires more work; we only mention the
result, we do not include a detailed proof here.)

There is also a sense in which stacks can be smooth.
For simplicity, consider only the notion of a topological
manifold, {\it i.e.}, a topological space that is locally
homeomorphic to ${\bf R}^n$ for some fixed $n$.
(Topological manifolds slightly generalize smooth manifolds;
 for simplicity, we shall only discuss topological manifolds here.)
A useful result, from which we can quickly see the
stack-theoretic definition, 
is that a topological space $N$ can be given the structure
of a topological manifold if and only if there exists 
another topological manifold $M$ (called an atlas) and a surjective
local homeomorphism $f: M \rightarrow N$.
Now, it is possible to generalize both ``surjective'' and
``local homeomorphism'' to stacks, so as the reader may guess,
the condition for a stack ${\cal F}$ to be a topological manifold is that
there exists a topological manifold $M$ (an honest space, not a stack),
together with a surjective local homeomorphism\footnote{
In the language we use here, a local homeomorphism is a continuous
representable map $f: X \rightarrow {\cal F}$ such that for all continuous
maps $g: Z \rightarrow {\cal F}$, the projection map
$\pi_1: Z \times_{ {\cal F} } X \rightarrow Z$ is a local homeomorphism.
An alternative and inequivalent 
definition is that near every point you have open substacks
$U$ and $V$ in the source and target so that the restricted map
is an isomorphism of stacks, {\it i.e.}, the map $f: U \rightarrow V$ is
representable, an equivalence of categories, and projection maps
between spaces appearing are local homeomorphisms.  Algebraic
geometers should also note that we are specifically {\it not} 
working in the Zariski
topology, so small open sets do exist, and the projection
map from a principal $G$-bundle to the base space, for $G$ finite,
is always a local homeomorphism. }
$M \rightarrow {\cal F}$.
The intuition behind the smoothness condition is that
the stack can be described locally with smooth coordinate patches.

For one last example, we shall outline how one can define a differential
form on a stack.  Again, the trick is to first rewrite the definition
in terms of pullbacks to other spaces, which can then be applied
to stacks.  A differential $n$-form $\omega$ on an ordinary smooth manifold $X$
is equivalent to the assignment, for each smooth manifold $Y$
and smooth map $f: Y \rightarrow X$, of a differential $n$-form
$\omega_f$ on $Y$ (thought of as $f^* \omega$),
subject to the condition that for each  
commuting diagram
\begin{displaymath}
\xymatrix{
Y_1 \ar[rr]^{g} \ar[dr] & & Y_2 \ar[dl] \\
 & X &
}
\end{displaymath}
(where $g$ is a smooth map), one has $g^* \omega_2 = \omega_1$.

As the reader has by now guessed,
to define a differential $n$-form on a smooth stack ${\cal F}$,
for each smooth manifold $Y$ and smooth map $f: Y \rightarrow {\cal F}$,
we associate a differential $n$-form $\omega_f$ on $Y$, such that
for each commuting diagram of the form above, $g^* \omega_2 = \omega_1$.

In particular, just as for functions, in the special case that
${\cal F}$ is a quotient stack $[X/G]$ (and ${\cal F}$ is smooth,
{\it i.e.}, $X$ is smooth and $G$ acts by diffeomorphisms), 
a differential $n$-form on 
${\cal F} = [X/G]$ is equivalent to a $G$-invariant differential form
on $X$.

Finally, we say a few words about the propogation of a worldsheet in a stacks,
as a special case of the above discussion.  A map
from the worldsheet $\Sigma$ to $[X/G]$ is specified by a diagram
\begin{displaymath}
\xymatrix{
{\cal P}  \ar[r]^{f} \ar[d]^{\pi} & X \\
\Sigma &
}
\end{displaymath}
where ${\cal P}$ is a principal $G$-bundle
$\Sigma$ and $f:{\cal P}\to X$ is $G$-equivariant.  
The action for a string sigma model on $[X/G]$ \cite{meqs} is formulated
on a lift of $\Sigma$ to ${\cal P}$ -- the resulting branch cuts
in the lift define the twisted sector.
Also note that possible twisted sectors are in one-to-one
correspondence with equivalence classes of principal $G$-bundles
${\cal P}$.
Note that we get an ordinary map $g:\Sigma\to X/G$ as follows.  
If $x\in \Sigma$,
pick any $p\in {\cal P}$ with $\pi(p)=x$.  Put $g(x)=\overline{f(p)}$, where
$\overline{f(p)}$ denotes the equivalence class in $X/G$ of $f(p)\in X$.  Note
that if $q\in{\cal P}$ also satisfies $\pi(q)=x$, then $q=g\cdot p$ for
some $g\in G$ and then equivariance of $f$ implies that 
\[
\overline{f(q)}=\overline{f(g\cdot p)}=\overline{g\cdot f(p)}=\overline{f(p)}
\]
so that $g$ is well defined.

In other words, we recover the usual notion of a worldsheet in an orbifold,
but now we can see additional structure over the singularities.  
More explicitly, vertex operators such as 
\begin{displaymath}
b^{\alpha \beta j_1 \cdots j_m}_{ 
\overline{\imath}_1 \cdots \overline{\imath}_n}(\phi_0)
\eta^{ \overline{\imath}_1 } \cdots
\eta^{ \overline{\imath}_n }
\theta_{j_1} \cdots \theta_{j_m}
\end{displaymath}
can be thought of as equivariant operators on ${\cal P}$ and identified
with equivariant sections of the appropriate bundle on a subspace of $X$.
Since $G$
can act nontrivially on the various fields used to construct the vertex
operator, we see that the formalism of stacks nicely encodes (and
generalizes) the more
familiar orbifold formalism, including twisted sectors, {\it etc}.

\section{Proofs of spectral sequences}  \label{pfs}

The existence of the spectral sequences used in the text
is essentially a straightforward consequence of the 
proofs of the analogues used in \cite{orig}.
Assume that $X$ is any smooth complex variety and $G$ is any complex 
reductive group acting on $X$.  Consider any two sheaves $A, B$ on $X$ 
and any spectral sequence which converges to $\mbox{Ext}^{\bullet}_{X}(A,B)$ 
and whose $E_{2}$ term consists of Hom's in the derived category 
$D^{b}(X)$, which are built functorially out of $A$ and $B$. (In our 
case $A = i_{*}E$ and $B = j_{*}F$.)  Now if the spectral sequence is 
functorial for pull-backs by automorphisms of $X$ (which our spectral 
sequences certainly are) and if we assume that $A$ and $B$ are 
$G$-equivariant it follows that $E_{2}^{p,q}$'s are $G$-modules and 
that all the differentials are maps of $G$-modules.  Since the 
invariants of a finite dimensional $G$-module split off canonically we 
get a spectral sequence of invariants:  $(E_{2}^{p,q})^{G}$ converging 
to $(\mbox{Ext}^{\bullet}_{X}(A,B))^{G}$.

An alternative method to prove the existence of the 
spectral sequences is to take a simplicial scheme
$(X/G)_{\bullet}$ presenting the stack $[X/G]$ and rewrite the proof 
of the existence of the spectral sequences from \cite{orig} 
on each layer of this 
simplicial scheme and then notice that the structure maps all respect 
the terms of the spectral sequence.  Combined with the fact that the 
augmentation map $(X/G)_{\bullet} \to [X/G]$ satisfies cohomological 
descent whenever $G$ is reductive (see {\it e.g.} 
\cite{sd} or \cite{deligne}) we 
now get the existence of the spectral sequence directly on the 
stack. To finish the prove we again have to invoke the fact that 
$\mbox{Ext}^{q}_{[X/G]}(A,B) = \mbox{Ext}^{q}_{Coh_{G}(X)}(A,B) = 
(\mbox{Ext}^{q}_{X}(A,B))^{G}$ both for $[X/G]$ and its closed substacks 
$[S/G]$ and $[T/G]$.

\end{document}